\documentclass[journal,draftcls,onecolumn,12pt,twoside]{IEEEtranTCOM}
\normalsize

\ifCLASSINFOpdf

\else

\fi

\usepackage[cmex10]{amsmath}

\usepackage[T1]{fontenc}
\usepackage{mathtools}

\DeclarePairedDelimiterX\Basics[1](){ #1}

\usepackage{amsmath,amssymb}

\ifCLASSINFOpdf
\usepackage[pdftex]{graphicx}

\else

\fi

\usepackage{amsmath}
\usepackage{stackengine}
\def\delequal{\mathrel{\ensurestackMath{\stackon[1pt]{=}{\scriptstyle\Delta}}}}
\usepackage{bm}
\usepackage{array}
\usepackage{tabu}
\usepackage{hyperref} 
\usepackage{graphics} 
\usepackage{algorithmic} 
\usepackage{algorithm}
\usepackage{bm}
\usepackage{graphicx}
\usepackage{subfigure}
\usepackage{subfloat}
\usepackage{amsmath}
\usepackage{caption}
\usepackage{amsthm,amssymb,amsmath}
\usepackage{lmodern}
\usepackage{mathrsfs}
\DeclareMathAlphabet{\mathpzc}{OT1}{pzc}{m}{it}
\usepackage{mathtools}

\usepackage{stackengine}
\def\delequal{\mathrel{\ensurestackMath{\stackon[1pt]{=}{\scriptstyle\Delta}}}}

\usepackage{amsmath}
\usepackage{cite}
\usepackage{graphicx}
\usepackage[font={small}]{caption}
\usepackage{ptext}

\usepackage{algorithm}
\usepackage{algorithmic}
\usepackage{float}

\usepackage{amsmath}
\usepackage{graphicx}
\usepackage{algorithmic}

\begin{document}

\title{Successive Wyner-Ziv Coding for the Binary CEO Problem under Logarithmic Loss}

\author{Mahdi~Nangir,
        Reza~Asvadi,~\IEEEmembership{Member,~IEEE,}
        ~Jun~Chen,~\IEEEmembership{Senior Member,~IEEE,}
        ~Mahmoud~Ahmadian-Attari,~\IEEEmembership{Member,~IEEE,}
        and Tad~Matsumoto,~\IEEEmembership{Fellow,~IEEE}
\thanks{M. Nangir was with the Faculty of Electrical Engineering, K.N. Toosi University of Technology, Tehran, Iran. He is now with the Faculty of Electrical and Computer Engineering, University of Tabriz, Tabriz, Iran. (e-mail: nangir@tabrizu.ac.ir).}
\thanks{R. Asvadi is with the Faculty of Electrical Engineering, Shahid Beheshti University, Tehran, Iran. (e-mail: r\_asvadi@sbu.ac.ir(corresponding author)).}
\thanks{J. Chen is with the Department of Electrical and Computer Engineering, McMaster University, Hamilton, ON, Canada. (e-mail: junchen@mail.ece.mcmaster.ca).}
\thanks{M. Ahmadian-Attari is with the Faculty of Electrical Engineering, K.N. Toosi University of Technology, Tehran, Iran. (e-mail: mahmoud@eetd.kntu.ac.ir).}
\thanks{T. Matsumoto is with the school of Information Science, Japan
	Advanced  Institute  of  Science  and  Technology,  Ishikawa  923-1292,  Japan,
	and  also  with  the  Centre  for  Wireless  Communications,  University  of  Oulu,
	90014 Oulu, Finland  (e-mail:  matumoto@jaist.ac.jp).}
\thanks{This paper was presented in part at the 29th Biennial Symposium on Communications (BSC 2018), Toronto, Canada \cite{n18suc}.}}
\maketitle
\begin{abstract}
The $L$-link binary Chief Executive Officer (CEO) problem under logarithmic loss is investigated in this paper. 
A quantization splitting technique is applied to convert the problem under consideration to a $(2L-1)$-step successive Wyner-Ziv (WZ) problem, for which a practical coding scheme is proposed. In the proposed scheme, low-density generator-matrix (LDGM) codes are used for binary quantization while low-density parity-check (LDPC) codes are used for syndrome generation; the decoder performs successive decoding based on the received syndromes and produces a soft reconstruction of the remote source. The simulation results indicate that the rate-distortion performance of the proposed scheme can approach the  theoretical inner bound based on binary-symmetric test-channel models.
\end{abstract}
\IEEEpeerreviewmaketitle

\section{Introduction}

Multiterminal source coding is an important subject of network information theory. Research on this subject has yielded insights and techniques that are useful for a wide range of applications, including, among other things, cooperative communications \cite{he2018} distributed storage \cite{kong10}, and sensor networks \cite{DPG09}. A particular formulation of multiterminal source coding, known as the Chief Executive Officer (CEO) problem, has received significant attention \cite{BZV96}. In this problem, there are $L$ encoders (also called agents), which observe independently corrupted versions of a source; these encoders compress their respective observations and forward the compressed data separately to a central decoder (also called CEO), which then produces a (lossy) reconstruction of the target source. 

The quadratic Gaussian setting of the CEO problem has been studied extensively, for which the rate-distortion region is characterized completely in the scalar case \cite{VB97,oha98,PTR04,CZB04,oha05,WCW10} and partially in the vector case \cite{WC13,WC14}. Extending these results beyond the quadratic Gaussian setting turns out to be highly non-trivial; there are some results in \cite{VV15,arxiv1,arxiv2}. Indeed, even for many seemingly simple sources and distortion measures, the understanding of the relevant information-theoretic limits is rather limited. A remarkable exception is a somewhat under-appreciated distortion measure called logarithmic loss (log-loss). As shown by Courtade and Weissman \cite{CW14}, the rate-distortion region of the CEO problem under log-loss admits a single-letter characterization for arbitrary finite-alphabet sources and noisy observations. Different from the conventional distortion measures which are typically imposed on ``hard'' reconstructions defined over the given source alphabet, the reconstructions associated with  log-loss are ``soft''. Specifically, in the context of the CEO problem, the most favorable ``soft'' reconstruction
is essentially the \textit{a posteriori} distribution of the source given the compressed data received from the encoders (which is a sufficient statistic); it is more informative than its ``hard'' counterparts and more suitable for many downstream statistical inference tasks. 

Recent years have seen significant interests in a new paradigm of wireless communications called cloud-radio access network (C-RAN). It has been recognized that the information-theoretic and coding-theoretic aspect of C-RAN is closely related to that of the CEO problem under log-loss \cite{ZXYC16}. This intriguing connection greatly enriches the implication of the latter problem and provides further motivations for the relevant research.

A main contribution of the present paper is a practical coding scheme for the CEO problem under log-loss. 
We adopt a hierarchical approach by decomposing the CEO problem into a set of simpler problems which the existing coding techniques can be directly brought to bear upon and then combining these small pieces to find the solution to the original problem. Two most basic problems in information theory are point-to-point channel coding and (lossy) source coding (also known as quantization). It is well known that the fundamental limits of these two problems can be approached using graph-based codes (e.g., low-density parity-check (LDPC) codes for channel coding \cite{Gallager63} and low-density generator-matrix (LDGM) codes for (lossy) source coding \cite{WM10}) in conjunction with iterative message-passing algorithms (e.g., the sum-product (SP) algorithm for channel decoding \cite{Gallager63} and the bias-propagation (BiP) algorithm  for (lossy) source encoding \cite{FILL07, FF07}). These basic coding components can serve as the building blocks of more sophisticated schemes for the problems at the second level of the hierarchy. Notable examples include the Gelfand-Pinsker problem and the Wyner-Ziv problem, which are solved via proper combination of source codes and channel codes \cite{WM09, NAA17}. With these solutions in hand, one can then tackle the problems at the third level or even higher. From this perspective, our proposed scheme for the CEO problem can be interpreted as successive implementation of Wyner-Ziv coding. 

The conversion of the CEO problem to the Wyner-Ziv problem is realized using quantization splitting. The idea of quantization splitting is by no means new. Indeed, it has been applied to the multiterminal source coding problem \cite{CB08} and multiple description problem \cite{CTBH06} among others \cite{ZCWB07}, particularly in the quadratic Gaussian setting. However, to the best of our knowledge, the application of quantization splitting is mainly restricted in the theoretical domain as a conceptual apparatus, and its practical implementation has not been addressed in the literature, at least for the problem under consideration (namely, the CEO problem under log-loss). In this work we mainly focus on the setting where the source is binary-symmetric and is corrupted by independent Bernoulli noises. It is worth emphasizing that this simple setting captures the essential features of the CEO problem and the methodology underlying our proposed scheme is in fact broadly applicable.

The organization of this paper is as follows. The problem definition and the concept of quantization splitting are presented in Section II. The proposed scheme is described in Section III. Section IV contains some analytical and numerical results. We conclude the paper in Section V.

\begin{table*}[t!]
	\caption{LIST OF SYMBOLS USED IN THIS PAPER.}
	\label{ttt}
	\centering
	\vspace{-20pt}
	\begin{center}
		\scalebox{.95}{
		\begin{tabular}{|c|c|}
			\hline
			\textbf{Symbol} & \textbf{Description}\\
			\hline
			$q_j$ & Appearance probability of the binary representation of $j$ in the links' output \\
			\hline
			$Q_j$ & Appearance probability of the binary representation of $j$ in the links' output for a specific $x \in \mathbb{B}$ \\
			\hline
			$p_l$ & Bernoulli noise parameter in the $l$-th link \\
			\hline
			$Y_l^n$ & Binary noisy observation in the $l$-th link \\
			\hline
			$n$ & Block length \\
			\hline
			$d_l$ & Crossover probability of the binary symmetric test channel in the $l$-th link \\
			\hline
			$\epsilon$ & Deviation from theoretical bounds \\
			\hline
			$D_{\text{em}}$ & Empirical achieved value of distortion under the log-loss criterion  \\
			\hline
			$d_{\text{H}}$ & Hamming distance \\
			\hline
			$P_l$ & Hamming distance between remote source and quantized sequence in the $l$-th link \\
			\hline
			$\phi_i(\cdot)$ & Mapping from $\mathbb{B} ^{K_i}$ to $\mathbb{B}$ \\
			\hline
			$\mu_{\text{max}}$ & Minimum value of $\mu$ for the case of $R_{\text{th}}=0$ and $D_{\text{th}}=1$ \\
			\hline
			$L$ & Number of links \\
			\hline
			$k_i$ and $k'_i$ & Number of syndrome bits in compound LDGM-LDPC codes \\
			\hline
			$m_l$ & Number of variable nodes in LDGM code $\mathcal{C}_{W_l}$ of the $l$-th link \\
			\hline
			$M_i$ & Number of variable nodes in LDGM code $\mathcal{C}'_i$ of the $i$-th link \\
			\hline
			$N_l^n$ &  Observation noise in the $l$-th link \\
			\hline
			$d_l^*$ & Optimal target values of $d_l$ \\
			\hline
			$W_l$ & Output sequence of splitter in the $l$-th link \\
			\hline
			${\bf H}$ & Parity-check matrix \\
			\hline
			$U_l$ & Quantized sequence in the $l$-th link \\
			\hline
			$R_l$ & Rate in the $l$-th link \\
			\hline
			$X^n$ & Remote binary symmetric source ($n$-tuple)\\
			\hline
			$\mu$ & Slope of the tangent line to sum rate-distortion bound curve \\
			\hline
			${\hat{X}}^n$ & Soft reconstruction of the remote source \\
			\hline
			$\delta_l$ & Splitting parameter in the $l$-th link \\
			\hline
			$R_{\text{th}}$($D_{\text{th}}$) & Theoretical bound of Sum-rate (Distortion)  \\
			\hline
			$\sigma$ & Well-ordered permutation \\
			\hline
		\end{tabular}
	}
	\end{center}
\end{table*}


\section{The CEO Problem and Quantization Splitting}

\subsection{Notations}

Throughout this paper, the logarithm is to the base $2$. Random variables and their realizations are shown by capital letters and lowercase letters, respectively. Sets and alphabet set of random variables are depicted by calligraphic letters. Furthermore, matrices are shown by bold-faced letters. 
The binary entropy function is $h_b(x)=-x \log x -(1-x) \log (1-x)$, $\mathbb{B} \delequal \{0,1\}$, and $p*d=p(1-d)+(1-p)d$ shows the binary convolution of $p$ and $d$. The list of symbols used in the paper is represented in Table \ref{ttt}.

\subsection{System Model}

\begin{figure}[t]
	\begin{center}
		\centering
		\includegraphics[width=4.4in,height=2.5in]{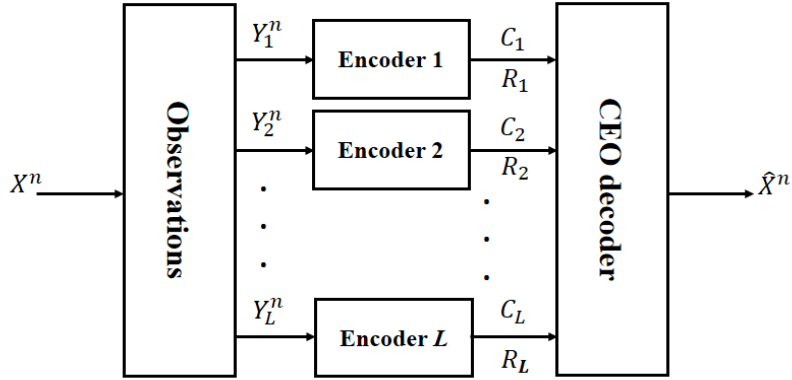}
	\end{center}
	\vspace{-20pt}
	\caption{Configuration of an $L$-link CEO problem.}
	\vspace{40pt}
	\label{CEO}
\end{figure}

Let $X^n=(X_1,X_2,\cdots,X_n)$ an independent and identically distributed (i.i.d.) remote source. $L$ noisy observations of $X^n$ are available in $L$ links that are mutually independent without any communication among them. These noisy observations, $Y^n_l$ for $l \in \mathcal{I}_L \delequal \{1,2,\cdots,L\}$, are generated by $X^n$ through independent memoryless channels.
The block diagram of an $L$-link CEO problem is depicted in Fig. \ref{CEO}. In each link, an encoder maps its noisy observation to a codeword $C_l$ by using a function $f_l$, as follows:
\begin{equation}
	\label{eqf}
	{C}_l=f_l ({Y}_l^n),\, \text{where}\;  {Y}_l^n \in \mathcal{Y}_l^n \; \text{and} \; {C}_l \in \mathcal{C}_l ,\, \text{for}\; l \in \mathcal{I}_L.
\end{equation}
The codewords $C_l$, for $l \in \mathcal{I}_L$, are sent to a joint CEO decoder via noiseless channels. The CEO decoder produces a soft reconstruction $\hat{X}^n=(\hat{X}_1,\hat{X}_2,\cdots,\hat{X}_n)$ of the original remote source $X^n$ by using a function $g$, as follows:
\begin{equation}
\label{eqg}
\hat{{X}}^n=g ({C}_1,{C}_2,\cdots,C_L), \, \text{where} \; ({C}_1,{C}_2,\cdots,C_L) \in \mathcal{C}_1 \times \mathcal{C}_2,\cdots \times \mathcal{C}_L.
\end{equation}
Specifically, $\hat{X}_t$ is decoder's approximation of the posterior distribution of $X_t$ given $(C_1,C_2,\cdots,C_L)$, $t\in\mathcal{I}_n$.

\textit{Definition 1:} The log-loss induced 
by a symbol $x\in\mathcal{X}$ and a probability distribution $\hat{x}$ on $\mathcal{X}$ is defined as
\begin{equation}
\label{eqlog}
d(x , \hat{x}) = \log \big({1 \over \hat{x} (x)}\big).
\end{equation}
More generally, for a sequence of symbols $x^n=(x_1,x_2,\cdots,x_n)$ and a sequence of distributions $\hat{x}^n=(\hat{x}_1,\hat{x}_2,\cdots,\hat{x}_n)$, let
\begin{equation}
\label{eqlogg}
d({x}^n , \hat{{x}}^n) ={1 \over n} \sum_{t=1}^{n} \log \big({1 \over \hat{x}_t (x_t)}\big).
\end{equation}

\textit{Definition 2:} A rate-distortion vector $(R_1,R_2,\cdots,R_L,D)$ is called strict-sense achievable under log-loss, if for all sufficiently large $n$, there exist functions $f_1$,$f_2$,...,$f_L$, and $g$ respectively according to (\ref{eqf}) and (\ref{eqg}) such that 
\begin{align}
\label{eqSTA}
R_l &\ge {1 \over n} \log \big |\mathcal {C}_l \big |, \, \text{for} \; l \in \mathcal{I}_L; \\ \nonumber
D &\ge \mathbb {E} (d({X}^n, \hat{{X}}^n)),
\end{align}
where $\mathbb {E}(\cdot)$ denotes expectation function. The closure of the set of all strict-sense achievable vectors $(R_1,R_2,\cdots,R_L,D)$ is called the  rate-distortion region of the CEO problem under log-loss and is denoted by $\overline {\mathcal {RD}}^{\star}_{\text{CEO}}$. 

\textit{Definition 3 (\cite[Definition 7]{CW14}):} Let ${\mathcal {RD}}^{i}_{\text{CEO}}$ be the set of all $(R_1,R_2,\cdots,R_L,D)$ satisfying
\begin{subequations}
\begin{equation}
\label{BTinner}
\sum_{l \in \mathcal{A}} R_l \ge I(Y_{\mathcal{A}};U_{\mathcal{A}} | U_{\mathcal{A}^c},Q),\quad \emptyset\subset\mathcal{A} \subseteq \mathcal{I}_L,
\end{equation}
\begin{equation}
\label{BTouterD}
D \ge H(X|U_{\mathcal{I}_L},Q),
\end{equation}
\end{subequations}
for some joint distribution
\begin{equation}
\label{key1}
p_Q(q)p_X(x) \prod_{l=1}^{L}p_{Y_l|X}(y_l|x)p_{U_l|Y_l,Q}(u_l|y_l,q),
\end{equation}
where in (\ref{BTinner}), $Y_{\mathcal{A}}=  \{Y_l:l \in \mathcal{A}\}$ and $\mathcal{A}^c=\mathcal{I}_L \backslash \mathcal{A}$. 

\textit{Definition 4 (\cite[Definition 8]{CW14}):} Let ${\mathcal {RD}}^{o}_{\text{CEO}}$ be the set of all $(R_1,R_2,\cdots,R_L,D)$ satisfying
\begin{equation}
\label{BTouter}
\sum_{l \in \mathcal{A}} R_l \ge [\sum_{l \in \mathcal{A}} I(Y_l;U_l | X,Q) +H(X |U_{\mathcal{A}^c},Q)-D]^+,\quad \emptyset\subset\mathcal{A} \subseteq \mathcal{I}_L,
\end{equation}
and (\ref{BTouterD}), for some joint distribution (\ref{key1}), where $[x]^+= \max \{0,x\}$ and $U_{\mathcal{A}} \leftrightarrow Y_{\mathcal{A}} \leftrightarrow X \leftrightarrow Y_{{\mathcal{A}}^c} \leftrightarrow U_{{\mathcal{A}}^c}$ forms a Markov chain for any $\mathcal{A} \subseteq \mathcal{I}_L$.


It is shown in \cite{CW14} that 
\begin{equation}
\label{eq:match}
\overline {\mathcal {RD}}^{\star}_{\text{CEO}}={\mathcal {RD}}^{i}_{\text{CEO}}={\mathcal {RD}}^{o}_{\text{CEO}};
\end{equation}
moreover, there is no loss of generality in imposing the cardinality bounds  $|\mathcal{U}_l| \le |\mathcal{Y}_l|, l\in\mathcal{I}_L$ and $|\mathcal{Q}| \le L+2$ on the alphabet sizes of auxiliary random variables $U_l$ and timesharing variable $Q$, respectively.

Given test channels $p_{U_l|Y_l}$, $l \in \mathcal{I}_L$, we define $\mathcal{RD}_{\text{CEO}}(p_{U_l|Y_l},l\in\mathcal{I}_L)$ to be the set of all $(R_1,R_2,\cdots,R_L,D)$ satisfying
\begin{align}
\label{BTouternQ}
\sum_{l \in \mathcal{A}} R_l &\ge I(Y_{\mathcal{A}};U_{\mathcal{A}} | U_{\mathcal{A}^c}),\quad \emptyset\subset\mathcal{A} \subseteq \mathcal{I}_L,\\
\label{BTouterDnQ}
D &\ge H(X|U_{\mathcal{I}_L}),
\end{align}
where $X$, $Y_{\mathcal{I}_L}$, and $U_{\mathcal{I}_L}$ are jointly distributed according to
$p_X(x) \prod_{l=1}^{L}p_{Y_l|X}(y_l|x)p_{U_l|Y_l}(u_l|y_l)$.

Note that (\ref{BTouternQ}) and (\ref{BTouterDnQ}) correspond respectively to  (\ref{BTinner}) and (\ref{BTouterD}) with timesharing variable $Q$ set to be a constant. Therefore, ${\mathcal {RD}}^{i}_{\text{CEO}}$ (as well as ${\mathcal {RD}}^{o}_{\text{CEO}}$ and $\overline {\mathcal {RD}}^{\star}_{\text{CEO}}$ in light of (\ref{eq:match})) can be expressed as the convex hull of the union of  $\mathcal{RD}_{\text{CEO}}(p_{U_l|Y_l}, l\in\mathcal{I}_L)$ over all $(p_{U_l|Y_l}, l\in\mathcal{I}_L)$. 
Moreover, we define $\mathcal{R}_{\text{CEO}}(p_{U_l|Y_l},l\in\mathcal{I}_L)$ to be the set of all $(R_1,R_2,\cdots,R_L)$ satisfying (\ref{BTouternQ}) and define its dominant face, denoted by $\mathcal{F}_{\text{CEO}}(p_{U_l|Y_l}, l\in\mathcal{I}_L)$, to be the set of $(R_1,R_2,\cdots,R_L) \in\mathcal{R}_{\text{CEO}}(p_{U_l|Y_l}, l\in\mathcal{I}_L)$ satisfying $\sum_{l=1}^L R_l = I(Y_{\mathcal{I}_L} ; U_{\mathcal{I}_L})$. Due to the contra-polymatroid structure of $\mathcal{R}_{\text{CEO}}(p_{U_l|Y_l}, l\in\mathcal{I}_L)$ \cite{CB08, ZCWB07}, $\mathcal{F}_{\text{CEO}}(p_{U_l|Y_l}, l\in\mathcal{I}_L)$ is non-empty and every $(R_1,\cdots,R_L,D)$ in $\mathcal{RD}_{\text{CEO}}(p_{U_l|Y_l}, l\in\mathcal{I}_L)$ is dominated, in a component-wise manner,  by $(R'_1,\cdots,R'_L,H(X|U_{\mathcal{I}_L}))$ for some $(R'_1,\cdots,R'_L)\in\mathcal{F}_{\text{CEO}}(p_{U_l|Y_l}, l\in\mathcal{I}_L)$. 

\subsection{Quantization Splitting}

$\mathcal{F}_{\text{CEO}}(p_{U_l|Y_l}, l\in\mathcal{I}_L)$ has $L!$ corner points. Specifically, each permutation $\pi$ on $\mathcal{I}_L$ is associated with a corner point $(R_1(\pi),\cdots,R_L(\pi))$ of $\mathcal{F}_{\text{CEO}}(p_{U_l|Y_l}, l\in\mathcal{I}_L)$ as follows:
\begin{align*}
R_{\pi(l)}(\pi)&=I(Y_{\pi(l)};U_{\pi(l)}|U_{\pi(l+1)},\cdots,U_{\pi(L)}),\quad l \in \mathcal{I}_{L-1},\\
R_{\pi(L)}(\pi)&=I(Y_{\pi(L)};U_{\pi(L)}).
\end{align*}
These corner points can be achieved via successive Wyner-Ziv coding with decoding order $U_{\pi(L)}\rightarrow U_{\pi(L-1)}\rightarrow\cdots\rightarrow U_{\pi(1)}$ (an implementation of this scheme for the case $L=2$ can be found in \cite{NAAC18}).

To achieve non-corner points of $\mathcal{F}_{\text{CEO}}(p_{U_l|Y_l}, l\in\mathcal{I}_L)$, we employ the quantization splitting technique introduced in \cite{CB08}, which is a generalization of the source splitting technique \cite{RU97} and a counterpart of the rate splitting technique in channel coding  \cite{RU96,GRU01}. Roughly speaking, the basic idea underlying the quantization splitting technique is that each non-corner point in the $L$-dimensional space can be projected to a corner point in the $(2L-1)$-dimensional space. Specifically, it is known \cite[Theorem 2.1]{CB08} that, for any rate tuple $(R_{1},R_{2}, \cdots , R_{L})\in\mathcal{F}_{\text{CEO}}(p_{U_l|Y_l}, l\in\mathcal{I}_L)$, there exist random variables $W_l$, $l \in \mathcal{I}_L$, and  a well-ordered permutation $\sigma$ \footnote{A well-ordered permutation is an arbitrary ordering of the set $\{W_1,W_2,\cdots,W_L,U_1,U_2,\cdots,U_L\}$ with $W_l$ appearing before $U_l$ for all $l \in \mathcal{I}_l.$} on the set $\{W_1,W_2,\cdots,W_{L},U_1,U_2,\cdots,U_L\}$ such that
\begin{align}
	\label{ee3}
	R_{l} &= I(Y_l;W_l|\{W_l\}^-_{\sigma})+I(Y_l;U_l|\{U_l\}^-_{\sigma}),\quad l\in\mathcal{I}_L,
\end{align}
where $\{W_{l}\}^-_{\sigma}$ and $\{U_{l}\}^-_{\sigma}$ represent the set of random variables that respectively appear before $W_{l}$ and $U_{l}$ in the well-ordered permutation $\sigma$; moreover, $W_l$ is a physically degraded version $U_l$, $l \in \mathcal{I}_L$, and at least one $W_l$ is independent of $U_l$ (and thus can be eliminated). 

It is instructive to view $U_l$ as a fine description of $Y_l$ and view $W_l$ as a coarse description split from $U_l$, $l \in \mathcal{I}_L$. Eq. (\ref{ee3}) suggests that the given rate tuple $(R_{1},R_{2}, \cdots , R_{L})$ can be achieved via successive Wyner-Ziv coding with decoding order specified by $\sigma$. It should be emphasized that the successive Wyner-Ziv coding scheme for non-corner points is in general more complicated than that for corner points. First of all, the scheme for non-corner points involves more encoding and decoding steps. Secondly and more importantly, to realize the splitting effect, one needs to generate a coarse-description codebook and then, for each of its codewords, generate a fine-description codebook; as a consequence, the number of finite-description codebooks grows exponentially with the codeword length,
causing a serious problem in practice. In this work we circumvent this problem by using a codebook construction technique inspired by the functional representation lemma \cite{WCZCP11, ELG}. Successive refinement coding scheme is also a multi-terminal encoding problem for, basically, downlink, where terminals are classified into several groups, each having different distortion requirements. The remote source is encoded such that the description for the groups having higher distortion requirement can help recover another groups having lower distortion 
requirement. Alternatively, our proposed coding scheme successively decodes binary observations and then softly reconstructs the remote source with a single value of distortion under the log-loss criterion.

\section{Description of the Proposed Scheme}

Consider an $L$-link binary CEO problem, where a remote binary-symmetric source (BSS) is corrupted by independent Bernoulli noises with parameters $p_1$, $p_2$, ... , and $p_L$, i.e.,
\begin{equation}
\label{ee1}
X \sim \text{Ber}({1 \over 2}),  \ \ \ \  Y_l=X \oplus N_l, \ \ \ \ N_l \sim \text{Ber}(p_l), \ \ \ \ l \in \mathcal{I}_L.
\end{equation}
We make the following two assumptions.
\begin{enumerate}
\item A binary-symmetric test channel model is adopted for each encoder. More specifically, it is assumed that $p_{U_l|Y_l}$ is a binary-symmetric channel (BSC) with crossover probability $d_l$, $l \in \mathcal{I}_L$. Hence, we can write $U_l = Y_l \oplus Z_l$, $l \in \mathcal{I}_L$, where $Z_l \sim \text{Ber}(d_l)$, $l \in \mathcal{I}_L$, are mutually independent and are independent of $(X,Y_{\mathcal{I}_L})$ as well. This assumption is justified by the numerical results in \cite{NAAC18}.

\item A BSC model is adopted for each splitter. More specifically, it is assumed that $p_{W_l|U_l}$ is a BSC with crossover probability $\delta_l$, $l \in \mathcal{I}_L$. Hence, we can write $W_l = U_l \oplus V_l$, $l \in \mathcal{I}_L$, where $V_l \sim \text{Ber}(\delta_l)$, $l \in \mathcal{I}_L$, are mutually independent and are independent of $(X,Y_{\mathcal{I}_L},U_{\mathcal{I}_L})$ as well. According to \cite[Definition 2]{GRU01}, this assumption incurs no loss of generality.
\end{enumerate}



Since the coding schemes associated with different well-ordered permutations  are conceptually similar, for ease of exposition, we focus on a specific permutation $\sigma=(W_1,W_2,\cdots,W_{L-1},$ $U_L,U_{L-1},\cdots,U_1)$ (we eliminate $W_L$ by setting $\delta_L=\frac{1}{2}$).
Each conditional mutual information in (\ref{ee3}) can be written as the difference of two terms, one associated with quantization and the other with binning.
As an example, consider the second term of $R_1$, i.e., $I(Y_1 ; U_1 |  W_1,\cdots,W_{L-1}, U_2,\cdots,U_L)$. We have
\begin{align}
&I(Y_1 ; U_1 |  W_1,\cdots,W_{L-1}, U_2,\cdots,U_L)\nonumber\\
&=I(Y_1;U_1|W_1,U_2,\cdots,U_L)\label{eq:degrade1}\\
&= I(U_2,\cdots,U_L,Y_1 ; U_1|  W_1) - I(U_2,\cdots,U_L ; U_1 |  W_1)\nonumber\\
&=I(Y_1 ; U_1|  W_1) - I(U_2,\cdots,U_L ; U_1 |  W_1),\label{eq:degrade2}
\end{align}
where (\ref{eq:degrade1}) is due to the degradeness of $W_l$ with respect to $U_l$, $l=2,\cdots,L-1$, and (\ref{eq:degrade2}) is because of the fact that $(U_1,W_1)$ and $(U_2,\cdots,U_L)$ are conditionally independent given $Y_1$.
The term $I(Y_1 ; U_1|  W_1)$ specifies the quantization rate needed to generate the fine description $U_1$ given the coarse description $W_1$ while the term $I(U_2,\cdots,U_L ; U_1 |  W_1)$ specifies the amount of rate reduction achievable through binning.


We use a binary quantizer to map outputs of a BSS to codewords of an LDGM code with a minimum Hamming distance. These quantizers are utilized in the encoders of our proposed coding scheme. Practically, binary quantization can be realized by using some iterative message passing algorithms such as the BiP algorithm \cite{FILL07} or the survey-propagation algorithm \cite{WM10}. Presence of side information can further reduce the compression rate required for a prescribed distortion constraint. Actually, this lossless source coding scenario can be practically realized by a binning operation based on channel coding schemes \cite{DPG09}.
In our proposed coding scheme, binning is implemented by using LDPC codes with the syndrome generation scheme.
This binning scheme is also used for the asymmetric Slepian-Wolf coding problem.
In practice, the SP algorithm can be used to iteratively decode the LDPC coset code specified by the given syndrome.

\subsection{The Proposed Coding Scheme: an Information-Theoretic Description}\label{sec:info}

To elucidate the overall structure of the proposed scheme, we first give a short description using the information-theoretic terminology. First, let  $W_L \delequal U_L$. In the following description, all the $\epsilon$ quantities are small positive real numbers.



\textbf{Codebook Generation:} 

\begin{enumerate}
	\item For $l \in \mathcal{I}_L$, construct a codebook $\mathcal{C}_{W_l}$ of rate $I(Y_l ; W_l)+\epsilon_{l,1}$ with each codeword generated independently according to $\prod_{t=1}^{n} p_{W_l}(\cdot)$.

	\item For $i \in \mathcal{I}_{L-1}$ and each codeword $w_i^n \in \mathcal{C}_{W_i}$, construct a codebook $\mathcal{C}_{U_i}(w_i^n)$ of rate $I(Y_i ; U_i | W_i)+\epsilon_{i,2}$ with each codeword generated independently according to $\prod_{t=1}^{n} p_{U_i|W_i}\big(\cdot |w_{i,t}\big)$.

	\item For $i=2,3,\cdots,L$, partition $\mathcal{C}_{W_i}$ into $2^{n[I(Y_i ; W_i |  W_1,\cdots,W_{i-1})+\epsilon_{i,3}]}$ bins, where each bin contains $2^{n[I(W_1,\cdots,W_{i-1} ; W_i)-\epsilon_{i,3}]}$ codewords.
	
	\item For $i \in \mathcal{I}_{L-1}$ and each codeword $w_i^n \in \mathcal{C}_{W_i}$, partition $\mathcal{C}_{U_i}(w_i^n)$ into $2^{n[I(Y_i ; U_i |  W_1,\cdots,W_i, U_{i+1},\cdots,U_L)+\epsilon_{i,4}]}$ bins, where each bin contains $2^{n[I(W_1,\cdots,W_{i-1}, U_{i+1},\cdots,U_L ; U_i |  W_i)-\epsilon_{i,4}]}$ codewords.
	
\end{enumerate}



\textbf{Encoding:}

\begin{enumerate}
		\item For $l \in \mathcal{I}_L$ and a given $y_l^n$, the $l$-th encoder finds a codeword $w_l^n \in \mathcal{C}_{W_l}$ that is jointly typical with $y_l^n$. Note that the Hamming distance between $w^n_l$ and $y^n_l$ is approximately $n(d_l * \delta_l)$.
		
		\item For $i \in \mathcal{I}_{L-1}$, the $i$-th encoder finds a codeword $u_i^n \in \mathcal{C}_{U_i}(w_i^n)$ that is jointly typical with $(y_i^n,w_i^n)$. Note that the Hamming distance between $u^n_i$ and $y^n_i$ is approximately $nd_i$ while the Hamming distance between $u^n_i$ and $w^n_i$ is approximately $n\delta_i$.
		
		\item For $l \in \mathcal{I}_L$, the $l$-th encoder sends the index $b(w_l^n)$ of the bin that contains $w_l^n$ (for $l=1$, it only sends the index $i(w_1^n)$ of $w_1^n$, and for $l=L$ nothing is sent), and the index $b(u_l^n)$ of the bin that contains $u_l^n$ to the decoder.
		
\end{enumerate}

\textbf{Decoding:}

\begin{enumerate}
	\item The decoder first decodes $w_1^n$ based on $i(w_1^n)$.
	
	\item For $i=2,\cdots,L$, it decodes $w_i^n$ by searching in the bin with index $b(w_i^n)$ for the unique codeword that is jointly typical with $(w_1^n,w_2^n,\cdots,w_{i-1}^n)$.
	
	\item For $j=L-1,\cdots,1$, it decodes $u_{j}^n$ by searching in the bin with index $b(u_{j}^n)$ for the unique codeword that is jointly typical with
	$(w_1^n,\cdots,w_{j}^n,u_{j+1}^n,\cdots,u_L^n)$.
	
	\item Finally, it uses $(\hat{u}^n_1,\cdots,\hat{u}^n_L)$ to produce a soft reconstruction of $x^n$ by the following rule:
	\begin{equation}
	\label{ss2}
	\hat{x}_t=p_{X|U_{\mathcal{I}_L}}(\cdot|\hat{u}_{1,t},\cdots,\hat{u}_{L,t}),
	 \quad t \in \mathcal{I}_n.
	\end{equation}
\end{enumerate}

\subsection{The Proposed Coding Scheme: a Coding-Theoretic Description}

Now we translate the above information-theoretic description of the proposed scheme to a coding-theoretic description. Along the way, we address certain practical issues encountered in codebook generation using a construction technique inspired by the functional representation lemma. For notional simplicity, the description is given for the case $L=3$; the extension to the general case is straightforward. 


\textbf{Codebook Generation:}

\begin{enumerate}
		\item For $l \in \mathcal{I}_3$, generate an LDGM codebook $\mathcal{C}_{W_l}$ with the rate of $I(Y_l;W_l)+\epsilon_{l,1}=1-h_b(d_l*\delta_l)+\epsilon_{l,1}$  \footnote{Note that $\delta_3=0$.} .
		
		\item For $i \in \mathcal{I}_{2}$ and each codeword $w_i^n$, construct a codebook $\mathcal{C}_{U_i}(w_i^n)$ as follows  \footnote{This construction is inspired by the functional representation lemma.} :
		
		Firstly, construct an LDGM code $\mathcal{C}_i^\prime$ with $2^{n[I(Y_i;U_i|W_i)+\epsilon_{i,2}]}$ codewords with each of length $n K_i$ where $K_i$ is a fixed integer. Let $\phi_i(\cdot)$ be a mapping \footnote{This is known as Gallager's mapping \cite{Gallager68}, which is widely used to construct source or channel codes with non-uniform empirical distribution \cite{SSCWW10, MHU18}.} from $\mathbb{B}^{K_i} \to \mathbb{B}$ such that 
		\begin{align}
		{|S_0| } \approx {2^{K_i}}(1-\delta_i),\quad {|S_1| } \approx {2^{K_i}} \delta_i,\label{ee66}
		\end{align}
		 where
	$S_b \delequal \{s^{K_i} \in \mathbb{B}^{K_i}: \ \phi_i(s^{K_i})=b\}$, $b\in \mathbb{B}$.
		Note that the approximation in (\ref{ee66}) can be made arbitrarily precise when ${K_i}\rightarrow\infty$. For each codeword $c^{n {K_i}} \delequal \big(c_1,c_2,\cdots,c_{n {K_i}}\big) \in \mathcal{C}_i^\prime$, map $c^{n {K_i}}$ to a codeword of length $n$ by using $\phi_i(\cdot)$ as below:
		\begin{equation}
		\label{ee7}
		\Big(\phi_i\big(c_1,\cdots,c_{K_i}\big),\phi_i\big(c_{{K_i}+1},\cdots,c_{2{K_i}}\big),\cdots,\phi_i\big(c_{(n-1) {K_i}+1},\cdots,c_{n {K_i}}\big)\Big).
		\end{equation}
		By doing this for all codewords in $\mathcal{C}_i^\prime$, a new codebook $\phi_i(\mathcal{C}_i^\prime)$ is obtained with $2^{n[I(Y_i;U_i|W_i)+\epsilon_{i,2}]}$ codewords, each of length $n$. Hence, the codebook $\mathcal{C}_{U_i}(w_i^n)$ can be defined as $w_i^n \oplus \phi_i(\mathcal{C}_{i}^\prime)$, which is a codebook obtained by adding $w_i^n$ to each codeword in $\phi_i(\mathcal{C}_{i}^\prime)$. Now consider the backward channels $Y_i = U_i \oplus Z_i^\prime$ and $U_i = W_i \oplus V_i^\prime$	\footnote{The representation of such backward channels can be viewed as a manifestation of the functional representation lemma. Moreover, it is  instructive to view $\phi_i(\mathcal{C}_{i}^\prime)$ as a codebook generated by $V_i^\prime$.}, where $Z_i^\prime\sim \text{Ber}(d_i)$, $ V_i^\prime\sim \text{Ber}(\delta_i)$, and $W_i$ are mutually independent. 		
	    It can be verified that
		\begin{align}
		\label{ee8}
		I(Y_i;U_i|W_i)=I(V_i^\prime \oplus Z_i^\prime;V_i^\prime)=h_b(\delta_i*d_i)-h_b(d_i).
		\end{align}
		
		\item For $i=2,3$, to partition $\mathcal{C}_{W_i}$ into $2^{n[I(Y_i;W_i|W_1,\cdots,W_{i-1})+\epsilon_{i,3}]}$ bins with each bin containing $2^{n[I(W_1,\cdots,W_{i-1};W_i)-\epsilon_{i,3}]}$ codewords, use an LDPC code of rate $I(W_1,\cdots,W_{i-1};W_i)-\epsilon_{i,3}$ with parity-check matrix ${\bf H}_i=(\tilde{\bf H}_i,\hat{\bf H}_i)$, where $\tilde{\bf H}_i$ is the parity-check matrix of $\mathcal{C}_{W_i}$. It can be verified that
		\begin{align}
		\label{ee10}
		I(Y_2;W_2|W_1)&=I(V_1^\prime \oplus Z_1^\prime \oplus N_1^\prime \oplus N_2; V_1^\prime \oplus Z_1^\prime \oplus N_1^\prime \oplus N_2 \oplus Z_2 \oplus V_2) \\ \nonumber
		&=H(V_1^\prime \oplus Z_1^\prime \oplus N_1^\prime \oplus N_2 \oplus Z_2 \oplus V_2) - H(Z_2 \oplus V_2) \\ \nonumber
		&=h_b(\delta_1 * d_1 * p_1 * p_2 * d_2 * \delta_2) - h_b(d_2 * \delta_2), \\ \nonumber
		I(W_1;W_2)&=1-H(V_1^\prime \oplus Z_1^\prime \oplus N_1^\prime \oplus N_2 \oplus Z_2 \oplus V_2) \\ \nonumber
		&=1-h_b(\delta_1 * d_1 * p_1 * p_2 * d_2 * \delta_2), \\ \nonumber
		I(W_1,W_2;U_3)&=H(W_1,W_2)-H(W_1,W_2|U_3)=1+h_b(\delta_1*d_1*p_1*p_2*d_2*\delta_2) \\ \nonumber
		&-H(Z_3^\prime \oplus N_3^\prime \oplus N_1 \oplus Z_1 \oplus V_1,Z_3^\prime \oplus N_3^\prime \oplus N_2 \oplus Z_2 \oplus V_2) \\ \nonumber
		&=H(W_1,W_2)-H(W_1,W_2,U_3)+1, \\ \nonumber
		I(Y_3;U_3|W_1,W_2)&=I(Y_3;U_3)-I(W_1,W_2;U_3)=1-h_b(d_3)-I(W_1,W_2;U_3).
		\end{align}
		
		\item For $i \in \mathcal{I}_{2}$, to partition $ \mathcal{C}^{\prime}_i $ into $ \ 2^{n[I(Y_i;U_i|W_1,\cdots,W_i,U_{i+1},\cdots,U_3)+\epsilon_{i,4}]} \ $ bins \footnote{This induces a partition of $ \ \mathcal{C}_{U_i}(w_i^n) \ $.}  with each bin containing $ 2^{n[I(W_1,\cdots,W_{i-1},U_{i+1},\cdots,U_3;U_i|W_i)-\epsilon_{i,4}]}$ codewords, we use an LDPC code with  $2^{n(I(W_1,\cdots,W_{i-1},U_{i+1},\cdots,U_3;U_i|W_i)-\epsilon_{i,4})}$ codewords, each of length $nK_i$.  The parity-check matrix of this LDPC code is ${\bf H}_1=(\tilde{\bf H}_1,\hat{\bf H}_1)$ for $i=1$ and ${\bf H}^{\prime}_2=(\tilde{\bf H}'_2,\hat{\bf H}'_2)$ for $i=2$, where $\tilde{\bf H}_1$ and $\tilde{\bf H}'_2$ are the parity-check matrices of $\mathcal{C}^{\prime}_1$ and $\mathcal{C}^{\prime}_2$, respectively. We have
		\begin{align}
		\label{ee9}
		I(U_2,U_3;U_1|W_1)&=I(V_1^\prime \oplus Z_1^\prime \oplus N_1^\prime \oplus N_2 \oplus Z_2 , V_1^\prime \oplus Z_1^\prime \oplus N_1^\prime \oplus N_3 \oplus Z_3 ; V_1^\prime)\\ \nonumber
		&=H(V_1^\prime \oplus Z_1^\prime \oplus N_1^\prime \oplus N_2 \oplus Z_2 , V_1^\prime \oplus Z_1^\prime \oplus N_1^\prime \oplus N_3 \oplus Z_3) \\ \nonumber
		&-H(Z_1^\prime \oplus N_1^\prime \oplus N_2 \oplus Z_2 , Z_1^\prime \oplus N_1^\prime \oplus N_3 \oplus Z_3), \\ \nonumber
		I(Y_1;U_1|W_1,U_2,U_3)&=I(Y_1;U_1|W_1)-I(U_2,U_3;U_1|W_1), \\ \nonumber
		I(W_1,U_3;U_2|W_2)&=I(V_2^\prime \oplus Z_2^\prime \oplus N_2^\prime \oplus N_1 \oplus Z_1 \oplus V_1, V_2^\prime \oplus Z_2^\prime \oplus N_2^\prime \oplus N_3 \oplus Z_3; V_2^\prime)\\ \nonumber
		&=H(V_2^\prime \oplus Z_2^\prime \oplus N_2^\prime \oplus N_1 \oplus Z_1 \oplus V_1, V_2^\prime \oplus Z_2^\prime \oplus N_2^\prime \oplus N_3 \oplus Z_3) \\ \nonumber
		&-H(Z_2^\prime \oplus N_2^\prime \oplus N_1 \oplus Z_1 \oplus V_1,Z_2^\prime \oplus N_2^\prime \oplus N_3 \oplus Z_3), \\ \nonumber
		I(Y_2;U_2|W_1,W_2,U_3)&=I(Y_2;U_2|W_2)-I(W_1,U_3;U_2|W_2).
		\end{align}
		
\end{enumerate}

\begin{figure}[t]
	\begin{center}
		\centering
		\includegraphics[width=5.2in,height=3in]{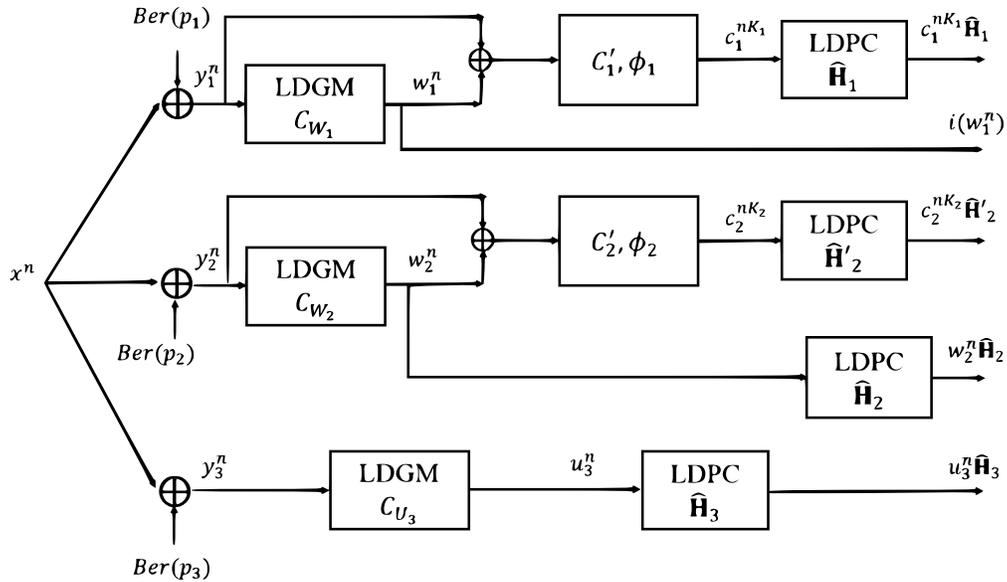}
	\end{center}
	\vspace{-15pt}
	\caption{The proposed encoding scheme.}
	\vspace{40pt}
	\label{ENC_op}
\end{figure}

\textbf{Encoding:} Different from the information-theoretic description in Section \ref{sec:info}, we shall interpret joint typicality encoding as minimum Hamming distance encoding, which is then implemented using the BiP algorithm.

\begin{enumerate}
	\item For $l \in \mathcal{I}_3$ and a given $y_l^n$, the $l$-th encoder finds a codeword $w_l^n \in \mathcal{C}_{W_l}$ from an LDGM code that is closest (in the Hamming distance) to $y_l^n$. 
	
	\item For $i \in \mathcal{I}_{2}$, find a codeword $c_i^{nK_i} \in \mathcal{C}_i^\prime$ such that $\phi_i(c_i^{nK_i})$ is the closest (in the Hamming distance) to $y_i^n \oplus w_i^n$. 
	
	\item Send the index of $w_1^n$ and the syndrome $c_1^{nK_1} \hat{{\bf H}}_1$ from the first link to the decoder; note that $w_1^n=i(w_1^n) {\bf G}_{W_1}$, where ${\bf G}_{W_1}$ is generator matrix of LDGM code $\mathcal{C}_{W_1}$. Also, send the syndromes $w_2^n \hat{{\bf H}}_2$ and $c_2^{nK_2} \hat{{\bf H}}^{\prime}_2$ from the second link to the decoder. Finally, send the syndrome $u_3^n \hat{{\bf H}}_3$ from the third link to the decoder.
	
\end{enumerate}

The block diagram of the proposed encoding scheme is depicted in Fig. \ref{ENC_op}.


\begin{figure}[t]
	\begin{center}
		\centering
		\includegraphics[width=5.2in,height=3.1in]{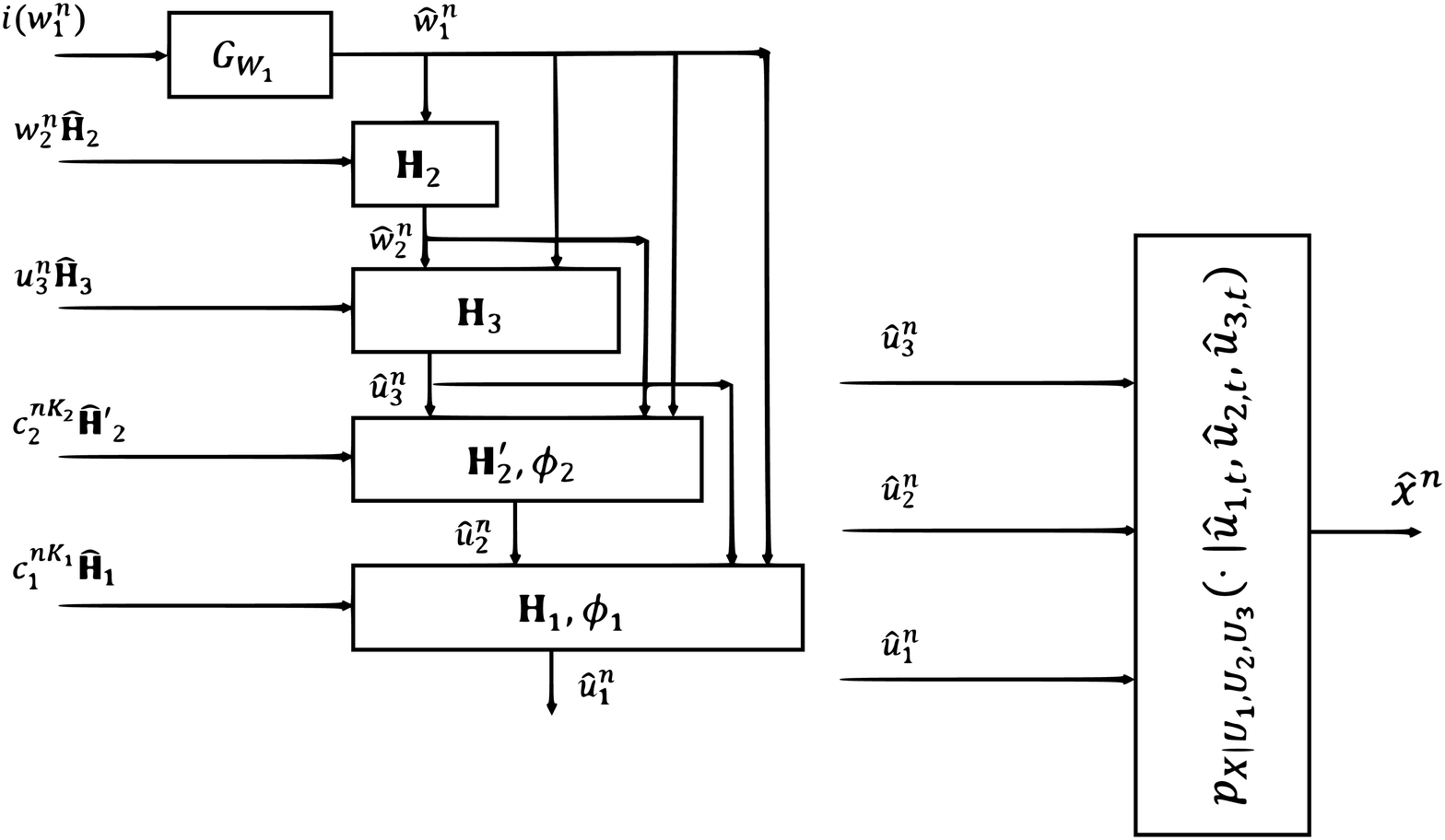}
	\end{center}
	\vspace{-15pt}
	\caption{The proposed successive decoding scheme.}
	\vspace{40pt}
	\label{DEC_op}
\end{figure}

\textbf{Decoding:} Different from the information-theoretic description in Section \ref{sec:info}, we shall interpret joint typicality decoding as maximum a posteriori decoding, which is then implemented using the  SP algorithm.

\begin{enumerate}
	\item The decoder first sets $\hat{w}_1^n=w^n_1$.
	\item It then finds the most likely choice of $w^n_2$, denoted by $\hat{w}_2^n$, based on $\hat{w}_1^n$ and $w_2^n {\bf H}_2$ (which can be deduced from $w_2^n \hat{{\bf H}}_2$ and the fact that $w_2^n \tilde{{\bf H}}_2$ is a zero vector). This can be realized via conventional Slepian-Wolf decoding with ${\bf H}_2$ defining the factor graph and $\hat{w}^n_1$ serving as side information (see, e.g., \cite{CHJ09}).
	
	\item It then finds the most likely choice of $u^n_3$, denoted by $\hat{u}_3^n$, based on $\hat{w}_1^n$, $\hat{w}_2^n$, and $u_3^n {\bf H}_3$ (which can be deduced from $u_3^n \hat{{\bf H}}_3$ and the fact that $u_2^n \tilde{{\bf H}}_3$ is a zero vector). This can be realized via conventional Slepian-Wolf decoding with ${\bf H}_3$ defining the factor graph and $(\hat{w}^n_1,\hat{w}^n_2)$ serving as side information. 
	
	\item It then finds the most likely choice of $c_2^{nK_2}$, denoted by $\hat{c}_2^{nK_2}$, based on $\hat{w}_1^n$, $\hat{w}_2^n$, $\hat{u}_3^n$, and $c_2^{nK_2} {\bf H}_2^\prime$ (which can be deduced from $c_2^{nK_2} \hat{{\bf H}}_2^\prime$ and the fact that $c_2^{nK_2} \tilde{{\bf H}}_2^\prime$ is a zero vector). This can be realized via joint demapping and decoding with $({\bf H}_2^\prime,\phi_2)$ defining the factor graph and $(\hat{w}_1^n,\hat{w}_2^n,\hat{u}_3^n)$ serving as the channel output 	
	(see, e.g., \cite{CHC13}). Set $\hat{u}^n_2=\hat{w}^n_2\oplus\phi_2(\hat{c}_2^{nK_2})$.

	\item It then finds the most likely $c_1^{nK_1}$, denoted by $\hat{c}_1^{nK_1}$, based on $\hat{w}_1^n$, $\hat{u}_2^n$, $\hat{u}_3^n$, and $c_1^{nK_1} {\bf H}_1$ (which can be deduced from $c_1^{nK_1} \hat{{\bf H}}_1$ and the fact that $c_1^{nK_1} \tilde{{\bf H}}_1$ is a zero vector). This can be realized via joint demapping and decoding with $({\bf H}_1,\phi_1)$ defining the factor graph and  $(\hat{w}_1^n, \hat{u}_2^n, \hat{u}_3^n)$ serving as the channel output. 
	Set $\hat{u}^n_1=\hat{w}^n_1\oplus\phi_1(\hat{c}_1^{nK_1})$.
	
    \item Finally, it produces a soft reconstruction $\hat{x}^n$ based on $\hat{u}_1^n$, $\hat{u}_2^n$, and $\hat{u}_3^n$ (see (\ref{ss2})).
\end{enumerate}

The block diagram  of the proposed decoding scheme is depicted in Fig. \ref{DEC_op}.

\subsection{Analysis of the Proposed Coding Scheme}

Now we proceed to specify the sizes of generator matrices and parity-check matrices used in the proposed scheme and other relevant parameters, assuming that $d_1$, $d_2$, $d_3$, $\delta_1$, and $\delta_2$ are given.


For the LDGM codes $\mathcal{C}_{W_l}$, shown in Fig. \ref{ENC_op}, their generator matrices are of size $m_l \times n$, $l \in \mathcal{I}_3$, respectively, where
\begin{align}
\label{e1}
{m_i \over n}&=1-h_b(d_i * \delta_i)+\epsilon_{i,1}, \ \ \ \ i \in \mathcal{I}_{2}, \\ \nonumber
{m_3 \over n}&=1-h_b(d_3)+\epsilon_{3,1}.
\end{align}
Furthermore, size of the generator matrix of the LDGM code $\mathcal{C}'_i$ is $M_i \times n K_i$, for $i \in \mathcal{I}_2$.
By properly designing these LDGM codes and increasing the block length $n$, one can ensure that
\begin{align}
\label{e18}
\mathbb{E} ({1 \over n} \sum_{j=1}^n [y_{i,j} \oplus w_{i,j}])&\approx d_i* \delta_i, \quad i \in \mathcal{I}_{2}, \\ \nonumber
\mathbb{E} ({1 \over n} \sum_{j=1}^n [y_{l,j} \oplus u_{l,j}])&\approx d_l,\quad 
\epsilon_{l,1}\approx 0,\quad l \in \mathcal{I}_3.
\end{align}






For the LDPC codes shown in Fig. \ref{ENC_op}, the sizes of their parity-check matrices are given as follows:
\begin{align}
\label{LDPC}
&{\bf H}_1: \ \ {n K_1 \times (n K_1 - M_1 +k_1)}, \ \ \ \ \ \ \ \ \ \ {\bf H}_2: \ \ {n \times (n-m_2+k_2)}, \\ \nonumber
&{\bf H}_2^\prime: \ \ {n K_2 \times (n K_2 - M_2 + k_2^\prime)}, \ \ \ \ \ \ \ \ \ \ {\bf H}_3: \ \ {n \times (n-m_3+k_3)}.
\end{align}
In the syndrome-decoding part of our proposed scheme, which is implemented by successive SP algorithms, if the optimized degree distributions for the BSC are used with sufficiently long LDPC codes, the bit error rate (BER) for the reconstruction of $\{U_1,U_2,U_3\}$ can be made very close to zero, i.e., $\text{BER}_l \approx 0$ for $l \in \mathcal{I}_3$. In such a case, the total distortion of the $l$-th link approximately equals $d_l$. In designing procedure of LDPC codes that are employed for the syndrome-generation and the syndrome-decoding, the following relations are considered in their code rates,

\begin{align}
\label{LDPCrates}
&{\bf H}_2: \  {m_2-k_2 \over n}=I(W_1;W_2)-\epsilon_{2,3}=1-h_b(P_1*\delta_1*P_2*\delta_2)-\epsilon_{2,3}, \\ \nonumber
&{\bf H}_3: \ {m_3-k_3 \over n}=I(W_1,W_2;U_3)-\epsilon_{3,3}=2+h_b(P_1*\delta_1*P_2*\delta_2)-H(W_1,W_2,U_3)-\epsilon_{3,3}, \\ \nonumber
&{\bf H}_2^\prime: \ {M_2-k_2^\prime \over n K_2}=I(W_1,U_3;U_2|W_2)-\epsilon_{2,4}=H(W_1,W_2,U_3)-H(W_1,U_2,U_3)-\epsilon_{2,4}, \\ \nonumber
&{\bf H}_1: \ {M_1-k_1 \over n K_1}=I(U_2,U_3;U_1|W_1)-\epsilon_{1,4}=H(W_1,U_2,U_3)-H(U_1,U_2,U_3)-\epsilon_{1,4},
\end{align}
where $P_l=p_l*d_l$ for $l \in \mathcal{I}_3$. Note that, there are four compound LDGM-LDPC codes \footnote{Generally, there is a compound LDGM-LDPC code in the first and the $L$-th link; and there are two compound codes in the $i$-th link, for $i=2,\cdots,L-1$. Thus, there are totally $2L-2$ compound codes in an $L$-link case.} in the proposed scheme for a $3$-link binary CEO problem. They comprise the LDGM codes $\mathcal{C}_{W_2}$, $\mathcal{C}_{U_3}$, $\mathcal{C}'_{1}$, and $\mathcal{C'}_{2}$, respectively with the LDPC codes having parity-check matrices ${\bf H}_2$, ${\bf H}_3$, ${\bf H}_1$, and ${\bf H}'_2$.


\section{Analytical Results}

It is clear that, for the proposed scheme, there is freedom in choosing $(d_1,\cdots,d_L)$ and $(\delta_1,\cdots,\delta_L)$. The role of $(d_1,\cdots,d_L)$ is to specify the dominant face $\mathcal{F}_{\text{CEO}}(p_{U_l|Y_l}, l\in\mathcal{I}_L)$ (and consequently the sum rate) while the role of $(\delta_1,\cdots,\delta_L)$ is to specify the location of the target rate tuple $(R_1,\cdots,R_L)$ on the dominant face. Note that for any $(R_1,R_2,\cdots,R_L,D)\in\mathcal{RD}_{\text{CEO}}(p_{U_l|Y_l}, l\in\mathcal{I}_L)$, we have
$\sum\limits_{l =1}^{L} R_l \ge R_{\text{th}}$ and $D \ge D_{\text{th}}$,
where
$R_{\text{th}}=I(Y_{\mathcal{I}_L};U_{\mathcal{I}_L})$, and $D_{\text{th}}=H(X|U_{\mathcal{I}_L})$.
One can interpret $R_{\text{th}}$ and $D_{\text{th}}$ as the minimum achievable sum rate and distortion associated with a given $(d_1,\cdots,d_L)$. Therefore, it is natural to choose $(d_1,\cdots,d_L)$ that achieves an optimal tradeoff between $R_{\text{th}}$ and $D_{\text{th}}$, which motivates the following definition.

\textit{Definition 5:} An $L$-tuple $(d_1^*,d_2^*,\cdots,d_L^*)$ is called an optimal $d$-allocation if it is a minimizer of $F$ for a certain $\mu \ge 0$, where
\begin{align}
\label{DR}
F= D_{\text{th}} + \mu R_{\text{th}}.
\end{align}


We shall derive several analytical results surrounding Definition 5. An investigation along this line was initiated in \cite{NAAC18} for the case $L=2$.

Note that
\begin{align*}
R_{\text{th}}& = H(U_{\mathcal{I}_L})-H(U_{\mathcal{I}_L}|Y_{\mathcal{I}_L})\\
&=H(U_{\mathcal{I}_L})-H(Z_{\mathcal{I}_L})\\
&=H(U_{\mathcal{I}_L})-\sum_{l=1}^L h_b(d_l),
\end{align*}
and 
\begin{align*}
D_{\text{th}}&=H(X,U_{\mathcal{I}_L})-H(U_{\mathcal{I}_L})\\
&=H(X)+H(U_{\mathcal{I}_L}|X)-H(U_{\mathcal{I}_L})\\
&=H(X)+H(N_{\mathcal{I}_L}\oplus Z_{\mathcal{I}_L})-H(U_{\mathcal{I}_L})\\
&=1+\sum\limits_{l=1}^L h_b(P_l)-H(U_{\mathcal{I}_L}),
\end{align*}
where $P_l=p_l*d_l$, $l \in \mathcal{I}_L$. 
Define
$Q_j=\prod\limits_{l=1}^L \eta(P_l, b_l(j))$ for $j=0,1,\cdots,2^L-1$,
where $b_l(j)$ denotes the $l$-th digit in the binary expansion of $j$, and
\begin{align*} 
\eta(P_l, b_l(j))=\left\{
  \begin{array}{ll}
    P_l, & b_l(j)=0, \\
    1-P_l, & b_l(j)=1.
  \end{array}
\right.
\end{align*}
For example, when $L=3$, we have
\begin{align}
	\label{P}
	Q_0 &=P_1P_2P_3,  &&Q_4 =(1-P_1)P_2P_3, \\ \nonumber
	Q_1 &=P_1P_2(1-P_3),   &&Q_5=(1-P_1)P_2(1-P_3), \\ \nonumber
	Q_2 &=P_1(1-P_2)P_3,  &&Q_6 =(1-P_1)(1-P_2)P_3, \\ \nonumber
	Q_3 &=P_1(1-P_2)(1-P_3),  &&Q_7 =(1-P_1)(1-P_2)(1-P_3).
\end{align}
It can be verified that
\begin{align}
\label{rec0}
H(U_{\mathcal{I}_L})=-\sum_{j=0}^{2^L-1}\big[{Q_j+Q_{2^L-1-j} \over 2}\big] \log \big[{Q_j+Q_{2^L-1-j} \over 2}\big].
\end{align}

\textit{Lemma 1:} For the objective function $F$ defined in (\ref{DR}), its minimum value is equal to 1 when $\mu\geq 1$.

\textit{Proof:} It is clear that $F=1$ when $(d_1,d_2,\cdots,d_L)=(0.5,0.5,\cdots,0.5)$. Now assume that a certain choice of $(d_1,d_2,\cdots,d_L)$ gives $F<1$. As a consequence, we have 
\begin{equation}
\label{mumu}
\mu<{1-D_{\text{th}} \over R_{\text{th}}}={H(U_{\mathcal{I}_L}) - \sum_{l=1}^{L}h_b(p_l*d_l) \over H(U_{\mathcal{I}_L}) - \sum_{l=1}^{L}h_b(d_l)} \le 1,
\end{equation}
which is contradictory with the fact that $ \mu\geq 1$. \qed

\textit{Lemma 2:} Let $p_1 \le p_2$ and $d_1 > d_2$. If $P_1=p_1*d_1$, $P_2=p_2*d_2$, $P'_1=p_1*d_2$, and $P'_2=p_2*d_1$, then
\begin{align}
\label{lem2}
P_1 + P_2 & > P'_1 + P'_2, \\ \nonumber
P_1  P_2 & > P'_1  P'_2, \\ \nonumber
2[P_1  P_2 - P'_1  P'_2] & = [P_1 + P_2] - [P'_1 + P'_2].
\end{align}

\textit{Proof:} The proof is straightforward.  \qed

\textit{Lemma 3:} Let $p_1 \le p_2 \le \cdots \le p_L$. If $d_1>d_2$ in the $L$-tuple $(d_1,d_2,\cdots,d_L)$, then by swapping $d_1$ and $d_2$, the value of $H(U_{\mathcal{I}_L})$ increases.

\textit{Proof:} See Appendix A.

\textit{Lemma 4:} If $p_1 \le p_2 \le \cdots \le p_L$, then $d_1^* \le d_2^* \le \cdots \le d_L^*$.

\textit{Proof:} Assume this is not true, and thus there exits $i$ such that $d_i^* > d_{i+1}^*$. We prove that by swapping $d_i^* $ and $ d_{i+1}^*$, the objective function $F=D_{\text{th}} + \mu R_{\text{th}}$ decreases which is a contradiction. Note that
\begin{align}
\label{obj}
F=1+\sum_{l=1}^{L} h_b(p_l*d_l^*) - \mu \sum_{l=1}^{L} h_b(d_l^*) + (\mu-1) H(U_{\mathcal{I}_L}).
\end{align}
Based on Lemmas $1$ and $2$, term $(\mu-1) H(U_{\mathcal{I}_L})$ decreases by swapping $d_i^* $ and $ d_{i+1}^*$. Also, the term $- \mu \sum_{l=1}^{L} h_b(d_l^*)$ clearly remains unchanged by this replacement. Without loss of generality, let us assume $i=1$. Therefore, it is enough to show that
\begin{align}
\label{ob}
h_b(p_1*d_2^*)+h_b(p_2*d_1^*) < h_b(p_1*d_1^*)+h_b(p_2*d_2^*).
\end{align}
By defining the following variables $z_1$ and $z_2$, we have:
\begin{align}
\label{obb}
& z_1 \delequal p_1*d_1^* \ \ \Rightarrow \ \ p_1*d_2^*<z_1<p_2*d_1^*, \\ \nonumber
& z_2 \delequal p_1*d_2^*+p_2*d_1^*-p_1*d_1^* \ \ \Rightarrow \ \ p_1*d_2^*<z_2<p_2*d_2^*<p_2*d_1^*, \\ \nonumber
& \Rightarrow \ \ z_1+z_2 = p_1*d_2^*+p_2*d_1^*.
\end{align}
Since, $h_b(x)$ is a concave function in $x$, from (\ref{obb})
\begin{align}
\label{obbb}
& h_b(p_1*d_2^*)+h_b(p_2*d_1^*) < h_b(z_1)+h_b(z_2).
\end{align}
Furthermore, $h_b(x)$ is an increasing function in the interval $[0,0.5]$. Thus,
\begin{align}
\label{obbb0}
& h_b(z_1)+h_b(z_2) < h_b(p_1*d_1^*)+h_b(p_2*d_2^*).
\end{align}
From (\ref{obbb}) and (\ref{obbb0}), the inequality (\ref{ob}) is concluded. Hence, the proof is completed. \qed




\section{Numerical Results}


Now we provide some numerical examples of optimal $d$-allocations.  Without loss of generality, we assume $p_1 \le p_2 \le \cdots \le p_L$. It follows by Lemma 4 that $d_1^* \le d_2^* \le \cdots \le d_L^*$ for the resulting optimal $d$-allocation. Obviously, $d_l^*$ equals $0$ for all $l$'s when $\mu=0$. There exists a $\mu_0 > 0$ such that for $0 \le \mu < \mu_0$, all the $L$ links are involved in information sending, i.e., $d_l^* < 0.5$ for $l \in \mathcal{I}_L$, while $d_L^*=0.5$ for $\mu=\mu_0$.
Therefore, the $L$-th link becomes inactive for $\mu \ge \mu_0$. Accordingly, the problem is reduced to an $(L-1)$-link case. By increasing $\mu$, the noisy links are eliminated one-by-one. Finally, it is reduced to the case of $L=2$. We illustrate this phenomenon through the following simple example.


\textit{Example 1:} Let $L=3$ and $p_1=p_2=p_3=0.1$. Based on the numerical results, if $0 \le \mu < \mu_0 \approx 0.3923$, then the straight line $0 \le d_1^*=d_2^*=d_3^*<0.125$ determines the location of the optimal points. For $\mu_0 \le \mu < \mu_1 \approx 0.42$, we have $d_1^*=d_2^* \le 0.125$ and $0.125<d_3^*<0.5$. If $\mu=\mu_1$, then $d_1^*=d_2^*=0.089$ and $d_3^*=0.5$. Similarly, if $\mu_1 < \mu < \mu_2 \approx 0.4245$, then $d_1^*<d_2^*<0.5$ and $d_3^*=0.5$. Next, for $\mu_2 \le \mu < \mu_{\text{max}}=0.64$, the first link is only involved in sending the information, i.e., $0.023<d_1^*<0.5$ and $d_2^*=d_3^*=0.5$. Finally, $d_1^*=d_2^*=d_3^*=0.5$ for $\mu \ge \mu_{\text{max}}$. 

The next example illustrates the sum-rate-distortion tradeoffs under equal $d$-allocation (i.e., $d_1=d_2=\cdots=d_L$). 

\begin{figure}[t]
	\centering
	\subfigure[{Equal noise parameters with the same quantizer in each link, the number of links is $L=3$.}]{\label{fig1}
		\includegraphics[width=3.1in,height=2.2in]{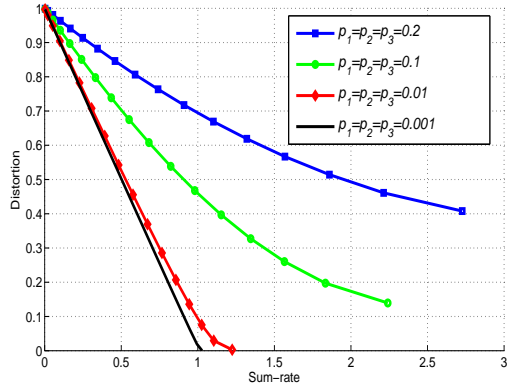}}
	\hspace{0.5mm}
	\subfigure[{$p_l=0.25$ for $l \in \mathcal{I}_L$.}]{\label{fig2}
		\includegraphics[width=3.1in,height=2.2in]{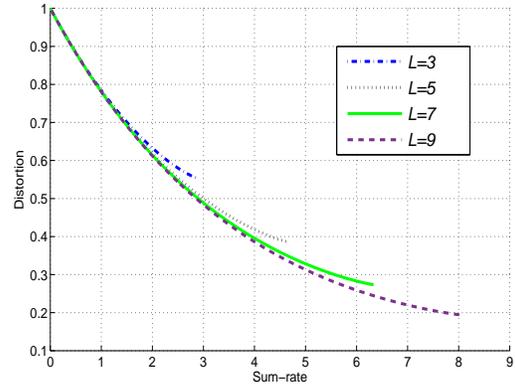}}
	\hspace{0.5mm}
	\vspace{-5pt}
	\caption{Sum-rate vs. distortion curves.}
	\vspace{40pt}
	\label{fig12}
\end{figure}

\textit{Example 2:} Let $L=3$ and $p_1=p_2=p_3$.  The sum-rate distortion curves under equal $d$-allocation are depicted in Fig. \ref{fig1} for various noise parameters. In Fig. \ref{fig2}, the sum-rate distortion curves under equal $d$-allocation are shown for the case of $p_l=0.25$ with $L=3,5,7,9$. 

\begin{figure}[t]
	\centering
	\subfigure[]{\label{fig3a}
		\includegraphics[width=3.1in,height=2.4in]{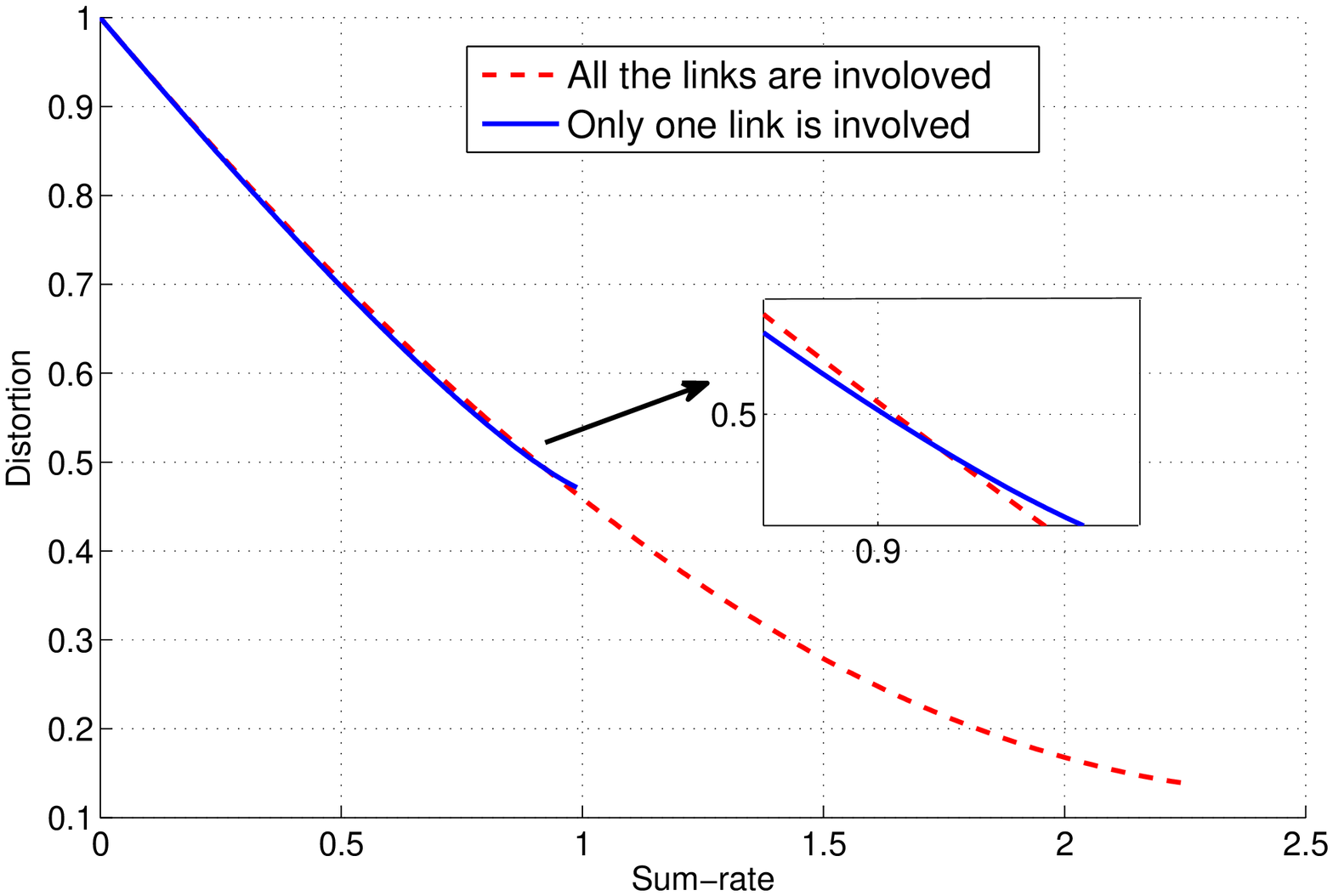}}
	\hspace{0.5mm}
	\subfigure[]{\label{fig3b}
		\includegraphics[width=3.1in,height=2.4in]{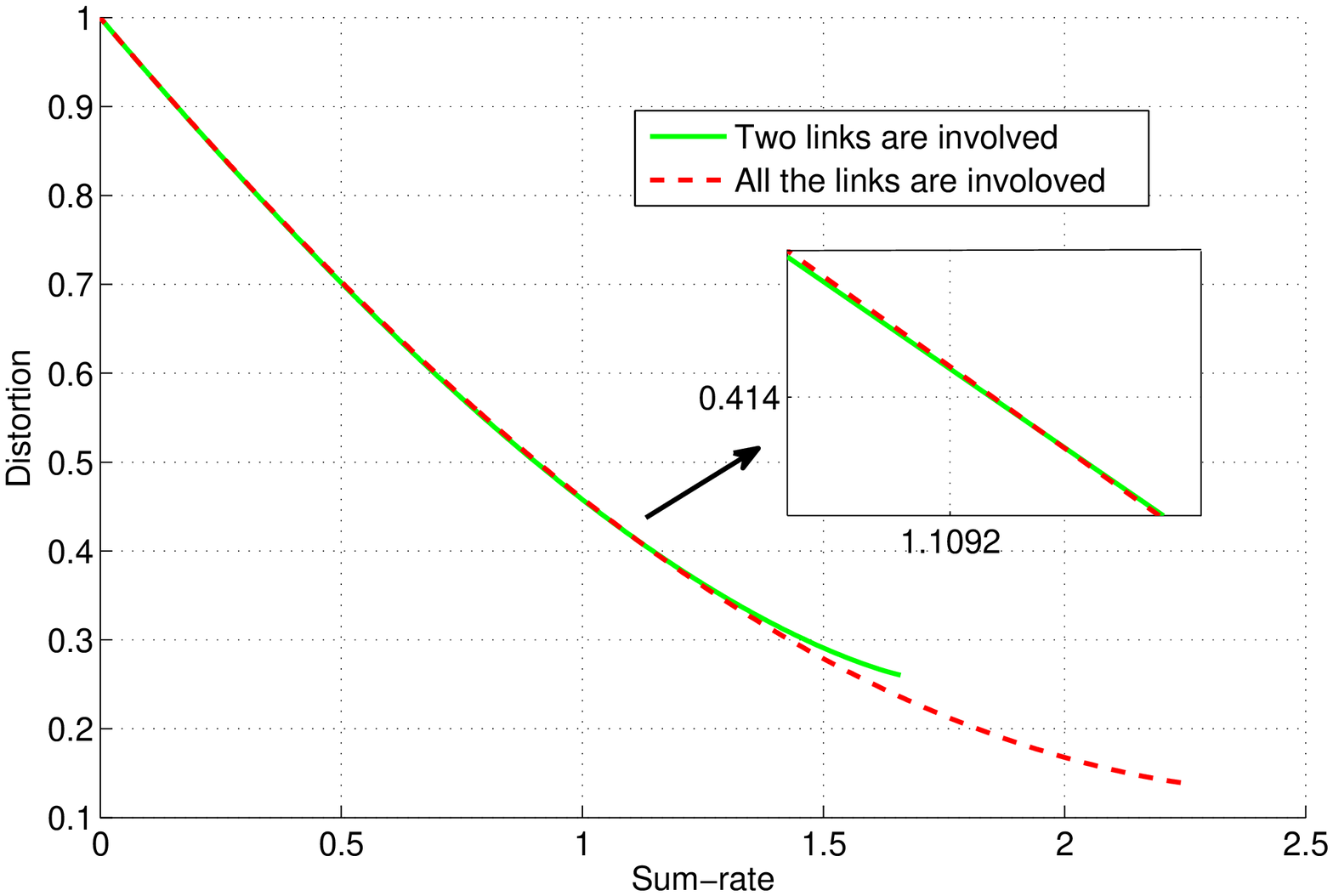}}
	\subfigure[]{\label{fig3c}
		\includegraphics[width=3.1in,height=2.4in]{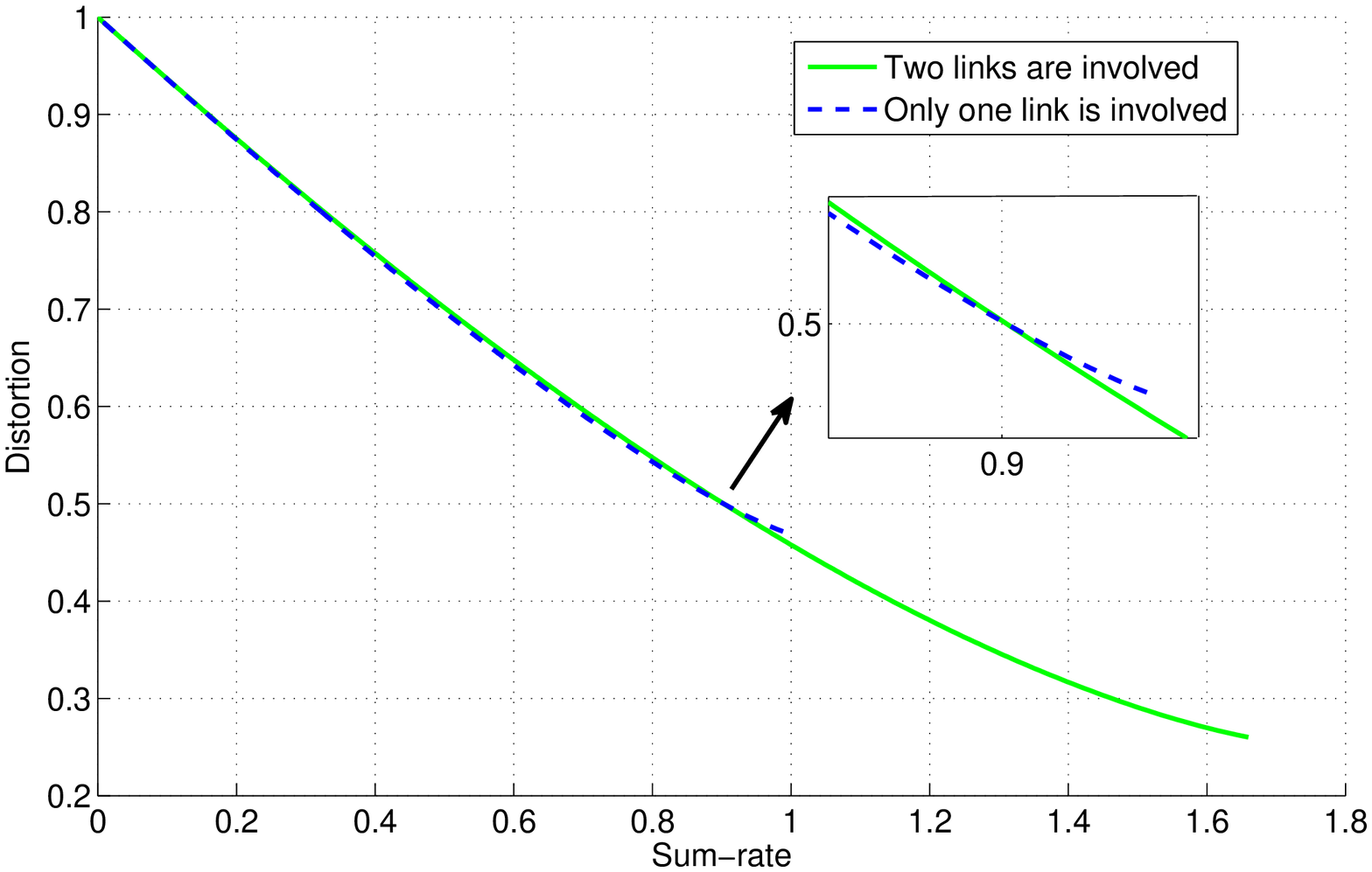}}
	\vspace{-10pt}
	\caption{Sum-rate vs. distortion curves, for $L=3$ and $p_1=p_2=p_3=0.1$ in different allocation scenarios.}
	\vspace{40pt}
	\label{fig3}
\end{figure}

\textit{Example 3:} Based on the numerical and the analytical results presented in \cite{NAAC18}, for a two-link binary CEO problem, the equal allocation, i.e., $d_1^*=d_2^*$, is not an optimal $d$-allocation for some values of sum-rate and distortion, even in the case of equal noise parameters $p_1=p_2$. Here, it is shown that this surprising result is also authentic for the multi-link case. In Fig. \ref{fig3}, the sum-rate distortion curves are shown for some cases. As it is seen, involving all the links does not necessarily provide minimum values of the sum-rate and the distortion.

\begin{figure}[t]
	\begin{center}
		\centering
		\includegraphics[width=4in,height=3in]{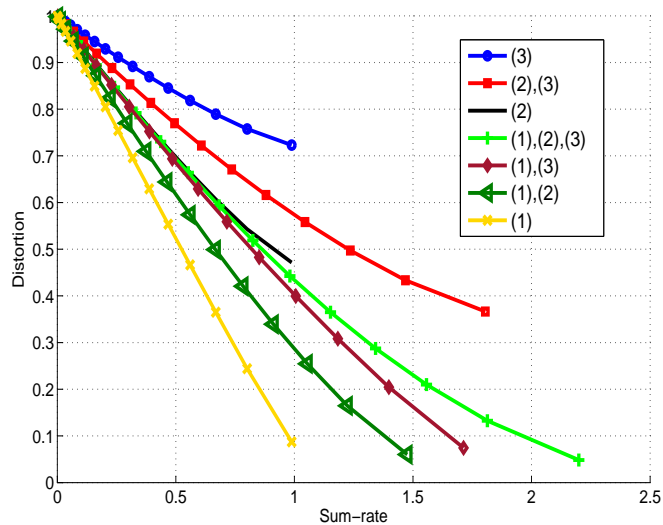}
	\end{center}
	\vspace{-10pt}
	\caption{\small{Sum-rate vs. distortion curves (Example $4$). The number of involved links are given in the legend.}}
	\vspace{40pt}
	\label{fig4}
\end{figure}

\textit{Example 4:} In this example, a $3$-link binary CEO problem is considered with almost prominent differences between values of the noise parameters. As an example, let $p_1=0.01$, $p_2=0.1$, and $p_3=0.2$. The sum-rate versus the distortion curves are presented in Fig. \ref{fig4}. It is assumed that the binary quantizers in each link are the same, when more than one link are involved in sending the information. Clearly, utilizing low noise links provides better results.



Now we proceed to present some experimental results for the proposed coding scheme. In our implementation, the degree distributions of the LDPC codes are provided in Appendix B; furthermore, the degree distributions of the LDGM codes are designed based on the method proposed in \cite{NAA17}, where the degrees of check nodes are regular and those of variable nodes follow a Poisson distribution. The relevant parameters of the proposed scheme are presented in Tables \ref{t2} and \ref{t1}. In particular, each choice of $(d_1,d_2,\cdots,d_L)$ corresponds to an optimal $d$-allocation. The rate of each encoder is calculated as follows:
\begin{align}
\label{rrr}
R_1& = \big({m_1 \over n}\big) + \big(I(Y_1;U_1 | W_1)\big) \times {k_1 \over n}  \\ \nonumber
& \approx \big(1-h_b(d_1*\delta_1)\big) + \big(h_b(d_1*\delta_1)-h_b(d_1)\big) \times {k_1 \over n} , \\ \nonumber
R_l& = \big({k_l \over n}\big) + \big(I(Y_l;U_l | W_l)\big) \times {k'_l \over n} \\ \nonumber
&= \big({m_l \over n}-{m_l-k_l \over n}\big)  + \big(I(Y_l;U_l | W_l)\big) \times {k'_l \over n} \\ \nonumber
& \approx \big(1-h_b(d_l*\delta_l)\big)-\big(I(W_1,\cdots,W_{l-1};W_l)\big) + \big(h_b(d_l*\delta_l)-h_b(d_l)\big) \times {k'_l \over n} ,\quad  2 \le l \le L-1, \\ \nonumber
R_L& = \big({k_L \over n}\big) = \big({m_L \over n}\big)- \big({m_L-k_L \over n}\big) \approx \big(1-h_b(d_L)\big) - \big(I(W_1,\cdots,W_{L-1};U_L)\big).
\end{align}

\begin{figure}[t]
	\begin{center}
		\centering
		\includegraphics[width=4in,height=3in]{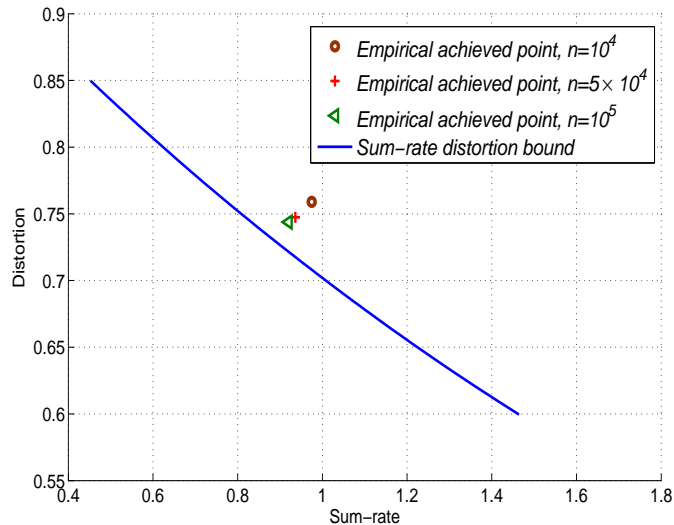}
	\end{center}
	\vspace{-15pt}
	\caption{\small{Performance of the sum-rate vs. the distortion for the implemented codes, (Example $5$).}}
	\vspace{40pt}
	\label{RD3}
\end{figure}

\textit{Example 5:} Consider a $3$-link case. Let $p_1=0.2$, $p_2=0.205$, and $p_3=0.21$ as well. For $\mu=0.25$, the optimal $d$-allocation is given by $d_1^*=0.1$, $d_2^*=0.164$, and $d_3^*=0.377$; as consequence, we have $R_{\text{th}}=0.9091$ and $D_{\text{th}}=0.7243$. The performance of the proposed coding scheme is presented for the corner and the intermediate points separately. The block lengths are equal to $n=10^4$, $n=5 \times 10^4$, and $n=10^5$. First, for achieving a corner point, there is no need to the splitter, i.e., $\delta_1=\delta_2=0$.
However, in order to achieve an intermediate point, any choice of $\delta_1 \in (0,0.5)$ and $\delta_2 \in (0,0.5)$ gives a specific intermediate point in the dominant face. In this example, we set $K_1=7$, $K_2=6$, $M_1=0.22 n K_1$, and $M_2=0.19 n K_2$ for the intermediate point. The results are presented in Table \ref{t2} and Fig. \ref{RD3}. The gap values for the code lengths $n=10^4$, $5 \times 10^4$, and $10^5$ are about $0.029$, $0.023$, and $0.02$, respectively.

\begin{table}[t]
	\caption{\footnotesize{PARAMETERS AND NUMERICAL RESULTS OF THE PROPOSED CODING SCHEME, (Example $5$).}}
	\label{t2}
	\centering
	\vspace{-20pt}
	\begin{center}
		\scalebox{0.7}{
			\begin{tabular} {| c | c | c | c | c | c | c | c |}
				\hline
				$m_1,m_2,m_3$ & $k_1,k_2,k'_2,k_3$ & $\delta_1,\delta_2$ & $d_1,d_2,d_3$ & $R_1,R_2,R_3,R$ & $\text{BER}_1$,$\text{BER}_2$,$\text{BER}_3$,$\text{BER}_4$ & $D_{\text{em}}$ &  $\text{Gap}$ \\
				\hline \hline
				$5400,4000,500$ & $-,3700,-,500$ & $0,0$ & $0.102,0.1669,0.3783$ & $0.54,0.4, 0.05,0.99$ & $0.0022,0.0025,-,-$ & $0.7532$ &  $0.0289$ \\
				\hline
				$3300,2100,500$ & $9900,2000,9950,500$ & $0.1024,0.141$ & $0.1036,0.1677,0.3783$ & $0.538,0.3868,0.05,0.9748$ & $0.001,0.0014,0.0014,0.0021$ & $0.759$ &  $0.0347$ \\
				\hline \hline
				$26500,18250,2300$ & $-,18000,-,2300$ & $0,0$ & $0.1014,0.1658,0.3778$ & $0.53,0.36,0.046,0.936$ & $0.0012,0.0.0016,-,-$ & $0.7474$ &  $0.0231$ \\
				\hline
				$16000,9500,2300$ & $49500,9500,49750,2300$ & $0.1024,0.141$ & $0.1019,0.166,0.3778$ & $0.5304,0.3677,0.0460,0.9441$ & $0.0009,0.0.0012,0.001,0.0015$ & $0.7552$ &  $0.0309$ \\
				\hline \hline
				$52500,36000,4500$ & $-,35000,-,4500$ & $0,0$ & $0.1009,0.1653,0.3776$ & $0.525,0.35,0.045,0.92$ & $0.001,0.0013,-,-$ & $0.7438$ &  $0.0195$ \\
				\hline
				$31500,18000,4500$ & $99000,17500,99500,4500$ & $0.1024,0.141$ & $0.1014,0.1648,0.3776$ & $0.5261,0.3542,0.045,0.9253$ & $0.0006,0.0011,0.0008,0.0014$ & $0.7503$ &  $0.026$ \\
				\hline
		\end{tabular}}
	\end{center}
\end{table}

\begin{figure}[t]
	\begin{center}
		\centering
		\includegraphics[width=4in,height=3in]{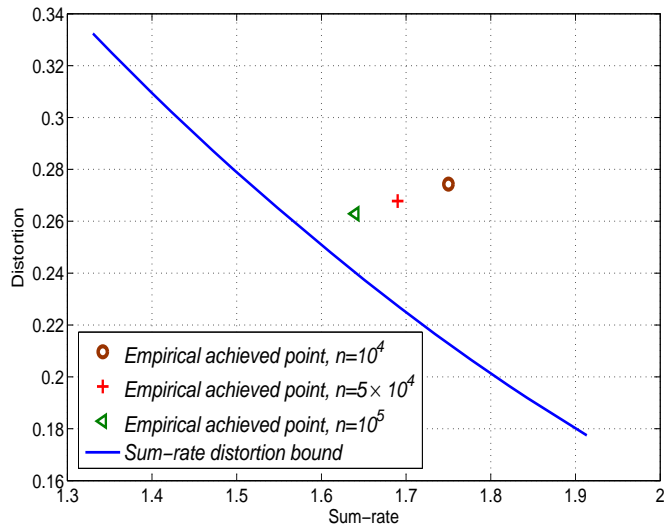}
	\end{center}
	\vspace{-15pt}
	\caption{\small{Performance of the sum-rate vs. distortion for the implemented codes, (Example $6$).}}
	\vspace{40pt}
	\label{RD4}
\end{figure}

\textit{Example 6:} Consider a $4$-link case and let $p_l=0.1$ for $l \in \mathcal{I}_4$. For $\mu=0.27$, the optimal  $d$-allocation  is given by $d^*_l=0.1$, for $l \in \mathcal{I}_4$; as a consequence, we have $R_{\text{th}}=1.591$ and $D_{\text{th}}=0.2534$.
The block lengths are assumed to be $n=10^4$, $n=5 \times 10^4$, and $n=10^5$. In order to achieve a corner point, there is no need to the splitter, i.e., $\delta_1=\delta_2=\delta_3=0$. However, for achieving an intermediate point, any choice of $\delta_i \in (0,0.5)$ for $i \in \mathcal{I}_3$, gives a specific intermediate point. 
In this example, we set $K_i=9$ and $M_i=0.12 n K_i$, $i \in \mathcal{I}_3$, for the intermediate point.
The results are shown in Table \ref{t1} and Fig. \ref{RD4}. The gap values for the code lengths $n=10^4$, $5 \times 10^4$, and $10^5$ are about $0.021$, $0.015$, and $0.01$, respectively.
According to the results of Examples $5$ and $6$, by decreasing the noise parameter or increasing the number of links, the gap values from the theoretical bounds are reduced. Moreover, larger block length $n$ causes smaller gap values.

\begin{table}[t]
	\caption{\footnotesize{PARAMETERS AND NUMERICAL RESULTS OF THE PROPOSED CODING SCHEME, (Example $6$).}}
	\label{t1}
	\centering
	\vspace{-20pt}
	\begin{center}
		\scalebox{0.6}{
			\begin{tabular} {| c | c | c | c | c | c | c | c |}
				\hline
				$m_{1\le i \le 3},m_4$ & $k_1,k_2,k'_2,k_3,k'_3,k_4$ & $\delta_1,\delta_2,\delta_3$ & $d_1,d_2,d_3,d_4$ & $R_1,R_2,R_3,R_4,R$ & $\text{BER}_1$,$\text{BER}_2$,$\text{BER}_3$,$\text{BER}_4$,$\text{BER}_5$,$\text{BER}_6$ & $D_{\text{em}}$ &  $\text{Gap}$ \\
				\hline \hline
				$5500,5500$ & $-,4400,-,4000,-,3600$ & $0,0,0$ & $0.1025,0.1028,0.1031,0.1022$ & $0.55,0.44,0.4,0.36,1.75$ & $0.0024,0.0027,0.003,-,-,-$ & $0.2743$ &  $0.0209$ \\
				\hline
				$5200,5500$ & $9950,4300,9950,3800,9950,3700$ & $0.01,0.01,0.01$ & $0.103,0.1029,0.1033,0.1022$ & $0.5441,0.4542,0.4041,0.37,1.7724$ & $0.0015,0.0019,0.0019,0.0026,0.0026,0.0022$ & $0.2754$ &  $0.022$ \\
				\hline \hline
				$27000,27000$ & $-,21500,-,19000,-,17000$ & $0,0,0$ & $0.102,0.1021,0.1025,0.1019$ & $0.54,0.43,0.38,0.34,1.69$ & $0.002,0.002,0.0023,-,-,-$ & $0.2678$ &  $0.0144$ \\
				\hline
				$25500,27000$ & $49600,20500,49600,18000,49600,17000$ & $0.01,0.01,0.01$ & $0.1019,0.1023,0.1025,0.1019$ & $0.5343,0.4342,0.3842,0.34,1.6927$ & $0.0016,0.0014,0.002,0.0022,0.0015,0.0018$ & $0.2694$ &  $0.016$ \\
				\hline \hline
				$53000,53000$ & $-,42000,-,37000,-,32000$ & $0,0,0$ & $0.1017,0.1019,0.102,0.1017$ & $0.53,0.42,0.37,0.32,1.64$ & $0.0011,0.0015,0.0012,-,-,-$ & $0.2629$ &  $0.0095$ \\
				\hline
				$50500,53000$ & $99100,40000,99100,34000,99100,32000$ & $0.01,0.01,0.01$ & $0.1015,0.101,0.1014,0.1017$ & $0.5293,0.4244,0.3643,0.32,1.638$ & $0.0009,0.0012,0.001,0.0013,0.001,0.0015$ & $0.2637$ &  $0.0103$ \\
				\hline
		\end{tabular}}
	\end{center}
\end{table}

\section{Conclusion}

We have proposed a practical coding scheme for the binary CEO problem under the log-loss criterion based on the idea of quantization splitting. The underlying methodology is in fact quite general and is applicable to the non-binary case as well. It should be emphasized that, to implement the proposed scheme, one needs to first specify the test channel model for each encoder. In general, it is preferable for the system to operate in a mode that corresponds to a certain boundary point of the rate-distortion region. Identifying the boundary-attaining test channel models is an interesting research problem worthy of further investigation.


\section*{Appendix A}

\begin{center}
Proof of Lemma 3
\end{center}

Since $H(U_{\mathcal{I}_L})$ is a function of $P_l$ for $l \in \mathcal{I}_L$, we shall denote it by $H_P (P_1,P_2,\cdots,P_L)$. 
It suffices to show that
\begin{equation}
\label{mum1}
H_P (P'_1,P'_2,P_3,\cdots,P_L) > H_P (P_1,P_2,P_3,\cdots,P_L),
\end{equation}
where $P'_1=p_1*d_2$ and $P'_2=p_2*d_1$. From (\ref{rec0}),
\begin{align}
\label{recc}
&H_P (P_1,P_2,\cdots,P_L)=-\sum_{j=0}^{2^L-1} q_j \log (q_j),
\end{align}
where $q_j = {Q_j+Q_{2^L-1-j} \over 2}$. Hence, (\ref{mum1}) can be written as follows:
\begin{equation}
\label{mum2}
-\sum_{j=0}^{2^L-1} q'_j \log (q'_j) > -\sum_{j=0}^{2^L-1} q_j \log (q_j).
\end{equation}
Partition $(q_j)$'s and $(q'_j)$'s in some groups with four members as follows:
\begin{align}
\label{partition}
&q_a = {P_1 P_2 \Psi + (1-P_1)(1-P_2) \Psi' \over 2}, \ \ \ \ &q_b = {P_1 (1-P_2) \Psi + (1-P_1)P_2 \Psi' \over 2}, \\ \nonumber
&q_c = {(1-P_1) P_2 \Psi + P_1(1-P_2) \Psi' \over 2}, \ \ \ \ &q_d = {(1-P_1)(1-P_2) \Psi + P_1 P_2 \Psi' \over 2},
\end{align}
and
\begin{align}
\label{partition1}
&q'_a = {P'_1 P'_2 \Psi + (1-P'_1)(1-P'_2) \Psi' \over 2}, \ \ \ \ &q'_b = {P'_1 (1-P'_2) \Psi + (1-P'_1)P'_2 \Psi' \over 2}, \\ \nonumber
&q'_c = {(1-P'_1) P'_2 \Psi + P'_1(1-P'_2) \Psi' \over 2}, \ \ \ \ &q'_d = {(1-P'_1)(1-P'_2) \Psi + P'_1 P'_2 \Psi' \over 2},
\end{align}
where $\Psi$ is an arbitrary product of $P_i$ or $(1-P_i)$, for $i=3,\cdots,L$, and
\begin{equation}
\Psi' = {P_3 \cdots P_L \times (1-P_3) \cdots (1-P_L) \over \Psi}.
\end{equation}

Without loss of generality, it can be assumed that $\Psi \le \Psi'$. Therefore, $q_a > q_d$ and $q'_a > q'_d$. By applying Lemma 2,
\begin{align}
\label{wq}
2(q_a-q'_a) &=\Psi [P_1P_2-P'_1P'_2]+\Psi' [P_1P_2-P_1-P_2-P'_1P'_2+P'_1+P'_2] \\ \nonumber
& =\Psi [P_1P_2-P'_1P'_2]+\Psi' [0] > 0 \Rightarrow q_a > q'_a.
\end{align}
Similarly, we can show $q'_d > q_d$. Thus, $q_a > q'_a > q'_d > q_d$. Now consider two following cases:
\begin{enumerate}
	\item $P_1 \ge P_2$:
	\begin{align}
	& P_1 \ge P_2 \ \ \ \Rightarrow \ \ \ q_c \ge q_b, \ \ \ q'_c \ge q'_b. \\ \nonumber
	& 2(q_c-q'_c)=\Psi [P'_1 P'_2 - P'_2+P_2-P_1 P_2] + \Psi' [P'_1 P'_2 - P'_1+P_1-P_1 P_2] \\ \nonumber
	&> \Psi [P'_1 P'_2 - P'_2+P_2-P_1 P_2] + \Psi [P'_1 P'_2 - P'_1+P_1-P_1 P_2]=0 \ \ \Rightarrow \ \ q_c>q'_c.
	\end{align}
	Note that in this case, $P'_1 P'_2 - P'_1+P_1-P_1 P_2={P'_2-P'_1+P_1-P_2 \over 2} \ge 0$. Similarly, we can show $q'_b > q_b$. Thus, $q_c > q'_c > q'_b > q_b$.
	\item $P_1 < P_2$:
	\begin{align}
	& P_1 < P_2 \ \ \ \Rightarrow \ \ \ q_c < q_b, \ \ \ q'_c < q'_b. \\ \nonumber
	& 2(q_c-q'_c)=\Psi [P'_1 P'_2 - P'_2+P_2-P_1 P_2] + \Psi' [P'_1 P'_2 - P'_1+P_1-P_1 P_2] \\ \nonumber
	&< \Psi' [P'_1 P'_2 - P'_2+P_2-P_1 P_2] + \Psi' [P'_1 P'_2 - P'_1+P_1-P_1 P_2]=0 \ \ \Rightarrow \ \ q_c<q'_c.
	\end{align}
	Note that in this case, $P'_1 P'_2 - P'_2+P_2-P_1 P_2={P'_1-P'_2+P_2-P_1 \over 2} \ge 0$. Similarly, we can show $q_b > q'_b$. Thus, $q_b > q'_b > q'_c > q_c$.
\end{enumerate}
Finally, note that
\begin{equation}
q_a+q_b+q_c+q_d = q'_a+q'_b+q'_c+q'_d = {\Psi + \Psi' \over 2}.
\end{equation}
Due to the concavity of the function $f(x)=-x \log (x)$, it is concluded that
\begin{align}
-q_a \log (q_a)-q_b \log (q_b)& -q_c \log (q_c) -q_d \log (q_d) \\ \nonumber
&< -q'_a \log (q'_a)-q'_b \log (q'_b) -q'_c \log (q'_c) -q'_d \log (q'_d) .
\end{align}
By doing a summation over all possible values of $\Psi$ in the mentioned $4$-tuple groups, (\ref{mum1}) is proved. \qed

\section*{Appendix B}

\begin{center}
Degree Distributions
\end{center}

In example 5, the employed degree distribution of parity-check matrices are as follows,

${\bf H}_2$ : $\lambda (x)= 0.4145x + 0.1667x^2 + 0.0571x^4+0.0737x^5 + 0.0022x^8+0.0118x^9 + 0.0751x^{11} + 0.0575x^{19}+0.0063x^{26} + 0.0046x^{35} + 0.0171x^{43}+0.0443x^{62} + 0.051x^{82} + 0.0165x^{99}$, and $\rho(x)= 0.5x^2 + 0.5x^3$.

${\bf H}_3$ : $\lambda (x)= 0.2911 x + 0.19x^2 + 0.0408x^4+0.0874x^5 + 0.0074x^6 + 0.1125x^7+0.0925x^{15} + 0.0186x^{20} + 0.124x^{32}+0.016x^{39} + 0.02x^{44}$, and $\rho(x)= x^3$.

${\bf H}_1$ : $\lambda (x)= 0.41 x + 0.1724 x^{2} + 0.0995x^{4}+0.0546x^{5} + 0.0379x^{6} + 0.0312x^{10}+0.0288x^{14} + 0.0432x^{16} + 0.0217x^{20} + 0.0385x^{28} + 0.0375x^{50}+0.0023x^{52} + 0.0158x^{62} + 0.0066x^{71}$, and $\rho(x)= 0.4 x^2 + 0.6 x^3$.

${\bf H}'_2$ : $\lambda (x)= 0.3424x + 0.165x^2 + 0.12x^4+0.0191x^5 + 0.012x^6 + 0.1416x^{10}+0.0211x^{25} + 0.0202x^{26} + 0.0185x^{34}+0.0429x^{36} + 0.0133x^{38} + 0.0022x^{39}+0.0104x^{40} + 0.0704x^{99}$, and $\rho(x)= 0.5x^2 + 0.5x^4$.

In example 6, the employed degree distribution of parity-check matrices are as follows, 

${\bf H}_2$ : $\lambda (x)= 0.3585x + 0.1664x^2 +0.0487x^4 + 0.1205x^5 + 0.0006x^{6} + 0.04x^{10} + 0.0744x^{13}+0.0339x^{25} + 0.0076x^{30} + 0.0564x^{34}+0.0918x^{99}$, and $\rho(x)= x^3$.

${\bf H}_3$ : $\lambda (x)= 0.3151x + 0.1902x^2 + 0.0449x^4+0.1706x^6 + 0.1405x^{17} + 0.0082x^{37}+0.044x^{41} + 0.0863x^{66}$, and $\rho(x)=  0.5x^3 + 0.5x^4$.

${\bf H}_4$ : $\lambda (x)= 0.292x + 0.174x^2 + 0.0523x^4+0.0257x^5 + 0.122x^6 + 0.0218x^8+0.021x^{10} + 0.0322x^{14} + 0.1128x^{23}+ 0.0328x^{31} + 0.0274x^{44}+0.0048x^{53} + 0.0126x^{59} + 0.0681x^{99}$, and $\rho(x)=  x^4$.

${\bf H}_1$, ${\bf H}'_2$, and ${\bf H}'_3$ : $\lambda (x) =0.3037x+0.1731x^2+0.0671x^4+0.0123x^5 + 0.1341x^6 + 0.0314x^{12}+0.011x^{14} + 0.0257x^{16} + 0.091x^{19}+0.04x^{39} + 0.0117x^{51}+0.0189x^{57} + 0.0112x^{62} + 0.0684x^{76}$, and $\rho(x)= 0.4x^2 + 0.6x^4$.


\end{document}